\documentclass[preprint,5p,11pt,times,twocolumn]{elsarticle}

\usepackage{amssymb}
\usepackage{lineno}
\usepackage[bookmarks,bookmarksnumbered]{hyperref}
\hypersetup{
    colorlinks = true,
    linkcolor = blue,
    anchorcolor =red,
    citecolor = blue,
    filecolor = red,urlcolor = red,
    pdfauthor=author
}

\usepackage{booktabs}
\usepackage{multirow}

\usepackage{graphicx}
\usepackage{float}
\usepackage{subfigure}
\usepackage{subcaption}
\usepackage{makecell}
\usepackage[ruled,linesnumbered]{algorithm2e}
\usepackage{bm}
\usepackage{amsmath}

\journal{Journal of Cleaner Production}

\begin{document}
\captionsetup[figure]{labelfont={bf},labelsep=period,name={Fig.}}

\begin{frontmatter}
    \title{Distributed Satellites Dynamic Allocation for Grids with Time Windows: A Potential Game Approach}
    \author{Weiyi Yang}
    \author{Yingwu Chen}
    \author{Xiaolu Liu}
    \author{Jun Wen}
    \author{Lei He\corref{cor1}}
    \cortext[cor1]{corresponding author, E-mail address: helei@nudt.edu.cn (L. He).}
    \address{College of Systems Engineering, National University of Defense Technology, Changsha, Hunan 410073, China}
    \begin{abstract}
         The allocation of tasks to a large number of distributed satellites is a difficult problem owing to dynamic changes in massive tasks and the complex matching of tasks to satellites. To reduce the complexity of the problem, tasks that are geographically close can be divided into a predefined grid with a specific time window and processed together. The problem then becomes a dynamic grid with time-window allocation problem (DGAP). To ensure consistent visibility between satellites and grids, the timeline of the DGAP is partitioned into several decision-making stages that are determined by dynamic changes in the time window. Subsequently, the DGAP can be resolved progressively adopting the potential game approach in the single-stage DGAP (sDGAP). First, to solve the discontinuity in the goal of the sDGAP, we approximate the goal by a smooth exponential sum function that we regard as the global utility function. Second, a potential game theoretic framework is constructed by decomposing this global utility function into the local utility functions of individuals. We prove that each Nash equilibrium of the proposed potential game is the optimal solution of the sDGAP. Third, to solve the potential game, a distributed algorithm, referred to as the selective time-variant better reply process (SeTVBRP) algorithm, is proposed and its convergence is proved. The SeTVBRP algorithm is an improved algorithm based on the better reply process algorithm, where two improvement methods (i.e., the selective action method and time-variant parameter method) are introduced. Through factor analysis, we demonstrate the effectiveness of the two improvement methods for the sDGAP. Last, numerical results show that the proposed algorithm outperforms existing learning algorithms and is effective in solving the DGAP.
    \end{abstract}
\end{frontmatter}


\section{Introduction}

Distributed satellites are platforms operated in different orbits in space and equipped with sensors to execute observation tasks \cite{She.2018}. An observation task refers to the imaging activity within a single switch of the equipped sensor. Driven by a rapid growth of observation tasks and enabled by the maturing of space technologies, the satellite field has developed from monolithic spacecraft to large-scale distributed satellites \cite{Wang.2019b}. For example, the company Planet has launched more than 150 Dove satellites in constructing a large-scale distributed satellite system \cite{Li.2022}. In addition to their obvious advantages, distributed satellites pose challenges relating to their control and cooperation. Allocating a massive observation task in a dynamic environment to satellites online is difficult, owing to the huge computing resource that is required \cite{Wang.2019}. There is thus a trend to make a grid target allocation of satellites \cite{An.2022}. Specifically, the observation tasks are divided into different grids according to geographical information. The computational complexity of the large-scale observation problem can be reduced and the time window information of the observation task can be preserved, providing a more accurate input for the satellite observation task planning in the grid.

The complexity of the dynamic allocation of grids with time windows mainly comes from the time-related property of satellite platforms and environment uncertainty, such as the limited observation time window, time-dependent observation quality and unpredictable changes in the grid observation load as the user requirement. Including the above properties and constraints, this paper focuses on the dynamic grid with a time winodow allocation problem (DGAP), where satellites are regarded as intelligent agents that make decisions adaptively. 

Due to considerations of robustness, scalability, and management overhead, centralized approaches designed for a single large satellite are not suitable for addressing problems involving distributed satellites in an uncertain environment \cite{Zheng.2019}. The concept of distributed self-organization comprehensively addresses the allocation of resources among tasks within a swarm of satellites in a distributed setting. Unlike the centralized allocation mechanism, the distributed allocation mechanism does not rely on a single main controller; instead, agents interact and communicate with each other to collectively achieve the global goal. However, a high-quality solution can hardly be guaranteed without the help of a centralized controller. Therefore, the core challenge in a distributed approach is to design an individual decision-making algorithm that contributes to the global goal theoretically. 

In this paper, the DGAP is formulated under the framework of the potential game proposed by Monderer and Shapley \cite{Marden.2009}. This framework addresses the challenge of distributed approaches by ensuring the perfect alignment of individual payoffs with global objectives. More importantly, the convergence of global optimal objectives can be guaranteed theoretically with the concept of the Nash equilibrium \cite{Li.2017}. 

Despite the advantages of applying potential game theory to the DGAP, the challenge of directly addressing the dynamic changes in the time window between the grid and satellite remains. To overcome this challenge, we segment the timeline of the DGAP into several continuous decision-making stages based on the changes in the time windows between satellites and grids. This ensures that the visibility relationship between the satellites and grids remains consistent within each stage, allowing us to effectively use the game approach for the single-stage DGAP (sDGAP). Consequently, the DGAP with multiple stages is resolved progressively by solving each individual stage.

The potential game approach can be divided into two hierarchical parts: the construction of the game framework and the design of distributed learning algorithms. However, the application of the potential game approach to the sDGAP still has difficulties. 
    
$\bullet$ The potential game theory mandates that the objective function of the problem be continuous. Nevertheless, the objective function of the sDGAP is discontinuous, rendering the direct application of the game framework infeasible. 
    
$\bullet$ The sDGAP involves multiple timing constraints, namely total allocation time constraints and satellite transition time constraints, and different satellite observation capacities for different grids. This makes the modeling of the problem into the potential game framework more complex. 

$\bullet$ The limited onboard computing and communication capacities of satellites make it difficult to apply general distributed learning algorithms of game theory to the sDGAP, which calls for the development of a novel and efficient distributed learning algorithm. Furthermore, the theoretical proof of the convergence of this algorithm presents difficulties.

The main contributions of the present paper are summarized as follows.

$\bullet$ In solving the discontinuity in the goal function of the sDGAP, we design a smooth global utility function to approximate the goal function and prove their equivalence. By decomposing the global utility function to each satellite, we construct a potential game model considering multiple complex timing constraints and varying satellite observation capacities.

$\bullet$ We propose a selective time variant better reply process (SeTVBRP) algorithm to solve the sDGAP. The convergence of our algorithm to the Nash equilibrium is proved.

$\bullet$ A numerical experiment is conducted to demonstrate the superiority of the SeTVBRP algorithm in terms of the solution quality and efficiency over state-of-the-art distributed learning algorithms. We publish the source codes of the algorithms and the datasets to enable future studies\footnote{The source code of our algorithm is available at https://github.com/yangweiyi15/satonlineallocation}

The remainder of this paper is organized as follows. Section 2 reviews approaches for solving the multi-satellite task allocation problem and clarifies the main contributions of the paper. Section 3 formulates the problem as a potential game model and proves the efficiency of the Nash equilibrium. Section 4 presents the SeTVBRP algorithm and a theoretical analysis of its convergence. Section 5 demonstrates the efficiency of new algorithmic features in numerical experiments and makes comparisons with recent state-of-the-art algorithms. The last section concludes the paper and gives a brief description of future work.

\section{Related works}
Approaches for solving the multi-satellite allocation problem can be divided into metaheuristic approaches, auction-based approaches and game theoretic approaches. Metaheuristic methods are popular for solving the multi-satellite allocation problem and are inspired by general phenomena and specific domain knowledge; e.g., genetic algorithms \cite{Kim.2015}, tabu search \cite{Sarkheyli.2013}, adaptive large neighborhood search \cite{He.2018} and simulated annealing \cite{Wu.2017}. Although metaheuristic approaches guarantee a high-quality solution, they are not suitable for the dynamic allocation of satellite systems owing to concerns relating to scalability and robustness. 

Auction-based approaches are another competitive means to deal with the distributed and dynamic scenario of this problem; e.g., the market-based approach \cite{Pica.2017}, contract network protocol \cite{Li.2020}, blackboard model approach \cite{Yang.2021} and consensus-based auction approach \cite{JohannesGerhardusvanderHorst.2012}. Yang et al. proposed an algorithm based on a blackboard model to coordinate the heterogeneous satellite system with the stochastic arrival of urgent tasks \cite{Yang.2021}. Horst et al. used a market-based mechanism to allocate tasks for a distributed satellite system and analyzed the general coordination mechanism of satellites in detail \cite{JohannesGerhardusvanderHorst.2012}. However, without the aid of a centralized controller, it is difficult for auction-based approaches to theoretically guarantee that the local decisions of the individual benefit the whole system, which limits the solution quality. 

In the past few years, game theoretic approaches have been shown to be a promising paradigm of distributed algorithm design for the multi-satellite allocation problem. Zheng et al. \cite{Zheng.2018} designed smoke signal play and broadcast-based play to allocate an emergency task, but they did not analyze the convergency of algorithms in depth. Sun et al. \cite{Sun.2021} proposed a dynamic scheduling approach based on a task merging policy, but they did not address the coupling issues of individual tasks. Wu et al. \cite{Wu.2020} formulated the distributed satellite task allocation problem under a potential game framework and proposed an algorithm with convergence analysis. Peng et al. \cite{Peng.2023} used a simplified version of the algorithm of Sun et al. \cite{Sun.2021} in addressing a sensor allocation problem. However, they only compared their algorithm with a basic market-based approach.

Our research is similar to the above studies and addresses multi-satellite task allocation using game theory. Two critical problems need to be solved: (i) most game theory studies have simplified the time-related property of satellite operation and few have considered the time windows of grids \cite{Sun.2018}; and (ii) the above models only consider the peer-to-peer matching of agents and tasks, and thus do not exactly match the heterogeneous workload of tasks and the observation times of satellites \cite{Wu.2020}. This paper attempts to solve these problems. To solve the first problem, we divide the timeline of the problem into multiple decision stages according to the change in the time window and adopt game theory approaches within a single stage. To solve the second problem, we enable a satellite to split the observation time and allocate the times to multiple grids in a decision-making stage, which realizes the efficient use of resources and more precise allocation.

\section{Problem Formulation}
The meanings of the main symbols and the abbreviations used in this paper are summarized in Table \ref{T0}.

\begin{table}[!ht]
    \centering
    \setlength{\belowcaptionskip}{0.2cm}
    \scriptsize
    \caption{Main symbols and notations}
    \begin{tabular}{ll}
    \hline
        Symbols & Meaning \\ \hline
        $S/{{s}_{i}}$ & Set of satellite/ $i$th satellite  \\ 
        $R/{{r}_{j}}$ & Set of grids in grid/ $j$th grid  \\ 
        $n/N$ & Number of satellites/ set of integers from 1 to $n$  \\ 
        $m/M$ & Number of grids in grid/ set of integers from 1 to $m$\\ 
        ${{w}_{ij}}$ & Time window of grid $r_j$ for satellite $s_i$   \\ 
        ${{c}_{ij}}$ & Beginning time of time window $w_{ij}$ \\ 
        ${{d}_{ij}}$ & End time of time window $w_{ij}$ \\ 
        $k$ & Index of decision-making stages  \\ 
        ${{t}_{k}}$ & Time point that state changes  \\ 
        ${{\beta }_{j}}/{{\beta }_{jk}}$ & Observation load of grid $r_{j}$ / Observation load of grid $r_{j}$ in stage $k$ \\ 
        ${{\alpha }_{ij}}/{{\alpha }_{ijk}}$ & observation capacity of satellite $s_{i}$ for grid $r_{j}$ / observation capacity \\ $/{{\alpha }_{j}}$& in stage $k$/ satellite observation capacity vector for grid $r_{j}$ \\ 
        ${{R}_{i}}/{{R}_{ik}}$ & Set of visible grids for satellite $s_{i}$ / set of visible grids in stage $k$ \\ 
        $R_{i}^{*}/R_{ik}^{*}$ & Set of allocated grids for satellite  / set of allocated grids in stage $k$ \\ 
        ${{S}_{j}}/{{S}_{jk}}$ & Set of visible satellites for $r_{j}$ / set of visible satellites in stage $k$ \\ 
        $S_{j}^{*}/S_{jk}^{*}$ & Set of allocated satellites for $r_{j}$ / set of allocated satellites \\ & for $r_{j}$ in stage $k$ \\ 
        ${{\eta }_{ik}}$ & Payload transfer time for satellite $s_{i}$ in stage $k$ \\ 
        ${{x}_{ij}}/x$ & Time that satellite $s_{i}$ allocated to grid $r_{j}$ / matrix formed by ${{x}_{ij}}$\\ 
        ${{y}_{j}}$ & Remaining observation load of grid $r_{j}$  \\ 
        ${{\rho }_{i}}$ & Imaging transition time for satellite $s_{i}$ \\ 
        $\varepsilon $ & Parameter in global utility function $U$  \\ 
        ${{A}_{i}}/{{a}_{i}}$ & Action set of satellite $s_{i}$ / allocation file of satellite $s_{i}$  \\ 
        ${{N}_{{{a}_{i}}}}$ & Set of grids that satellite $s_{i}$ allocates in action  $a_{i}$   \\ 
        $S_{a}^{j}$ & Satellite set allocated to grid $r_{j}$ in an action file $a$ \\ 
        $U(a)$ & Global utility function of action file $a$ / local utility function \\$/{{U}_{i}}(a)$& of action file $a$ of satellite $s_{i}$\\ 
        $t$ & Iteration index in the SeTVBRP\\ 
        ${{T}_{\max }}$ & Maximum value of $t$ in the SeTVBRP\\ 
        $\varepsilon (t)$ & Value of parameter $\varepsilon$ at iteration $t$ \\ 
        $\omega (t)$ & Value of parameter $\omega$ at iteration $t$\\ 
        $A_{i}^{t}$ & Selective action set of satellite $s_{i}$ at iteration $t$ \\ 
        $B_{i}^{t}$ & Better reply action set of satellite $s_{i}$ at iteration $t$ \\ 
        $\vartheta $ & Probability that $a$ player takes same action as in last iteration  \\ 
        ${{T}_{f}}$ & Minimum value of iteration $t$ that $\varepsilon (t)$ is minimized.  \\ 
        $a(t)/{{a}^{k}}(t)$ & Allocation file at iteration $t$ / allocation file at iteration $t$ in stage $k$   \\ 
        $\tau $ & Parameter that controls iteration when $\varepsilon (t)$ begin to decrease \\ 
        $\varphi $ & Parameter that controls iteration when $\omega (t)$ begin to decrease  \\ \hline
    \end{tabular}
    \label{T0}
\end{table}

\subsection{Problem description}
\subsubsection{Formulation of DGAP}

\begin{figure}
	\centering
    \includegraphics[scale=0.48]{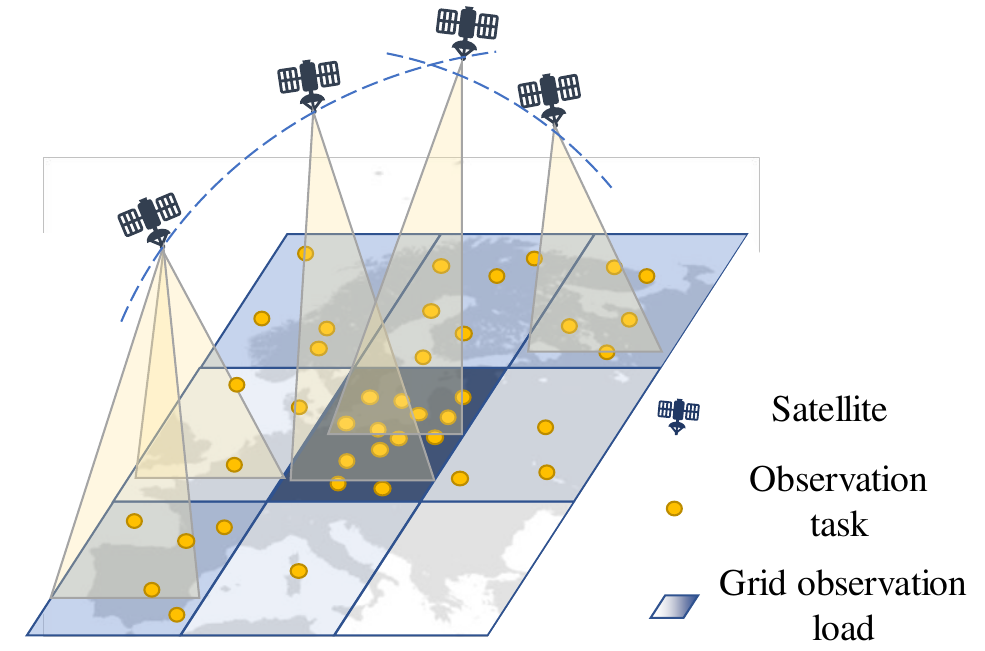}  
	\caption{Satellites and grids distribution diagram}   
\label{f1}
\end{figure}

The DGAP for distributed satellites refers to a set of satellites $S = \left\{ {{s_i}|i \in N} \right\}$, $N=\left\{ 1,2,...,n \right\}$,  and a set of grids $R=\left\{ {{r}_{j}}|j\in M \right\}$, $M=\left\{ 1,2,...,m \right\}$. As shown in Fig. \ref{f1}, a satellite can only perform tasks within one grid simultaneously, and a grid requires at least one satellite for observation. Each grid has an observation load that can dynamically change with time and is affected by the number the observation tasks directly. The yellow dots in the figure represent the observation tasks, and the depth of the grid color represents the grid’s observation load. The present problem assumes that real-time communication is accomplished via these inter-satellite links, thereby endowing the satellite system with distributed decision-making autonomy.

\begin{figure}[t]
	\centering
    \includegraphics[scale=0.28]{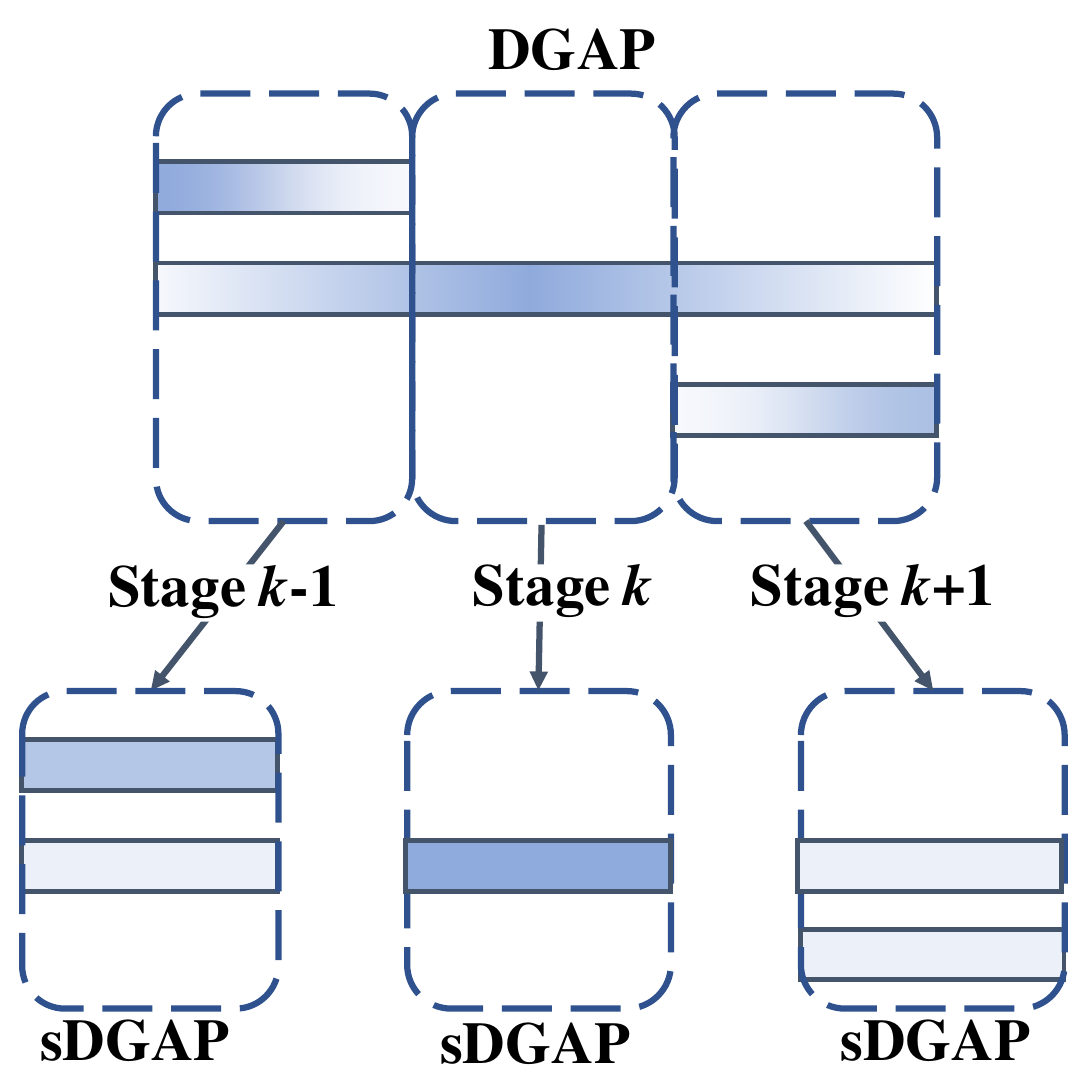}  
	  \caption{Schematic of the observation capabilities of different satellites for different grids}   
\label{f2}
\end{figure}

For each satellite $s_i$, the time window of grid $r_j$ is defined as ${{w}_{ij}}=[{{c}_{ij}},{{d}_{ij}}]$, where $c_{ij}$ is the time window beginning time and $d_{ij}$ is the time window end time.  The grid is fixed on the ground, and its time window can be precalculated based on the relative position of the satellite trajectory and grid. Changes in the time window primarily result from limitations in payload field-of-view, obstructions caused by cloud cover, suboptimal nighttime lighting conditions, and potential satellite malfunctions. We refer to the moment when the time window changes as the state change point $t_k$, and the time interval $[{{t}_{k}},{{t}_{k}}+\vartriangle t]$ as stage $k$. The timeline of the DGAP is divided into three decision-making stages based on the dynamic changes in the time window, as illustrated in the Fig. \ref{f2}. Therefore, a series of continuous single-stage DGAPs (sDGAPs) are formed by segmenting the timeline, with varying model parameters reflecting the dynamic characteristics of DGAP at each stage. The DGAP can be solved by sequentially solving these sDGAPs

For each stage $k$, the observation load of grid $r_j$ is denoted $\beta_{jk}$. As shown in the Fig. \ref{f2}, a time window can be divided into several decision-making stages, and the observation capacity can differ among these decision stages. For instance, the decision stages closer to the central period of a time window can have a larger observation capacity. We denote the observation capacity of satellite $s_i$ for grid $r_j$ by $\alpha_{ijk}$. 
    
According to the definition of state change moment, in stage $k$, the observation capacity $\alpha_{ijk}$ and the observation load $\beta_{jk}$ are constant. Let $R_{ik}$  represent the set of visible grids of satellite $s_i$ in stage $k$:
\begin{equation}
\begin{array}{l}
{{R}_{ik}}=\{{{r}_{j}}|{{t}_{k}}<{{c}_{ij}},{{t}_{k}}+\Delta t>{{d}_{ij}}\}
\end{array}
\label{eq1}  
\end{equation}

\begin{figure}[!htb]
    \centering
    \subfigure[Transition time ${{\eta }_{ik}}=H$]{
        \includegraphics[scale=0.48]{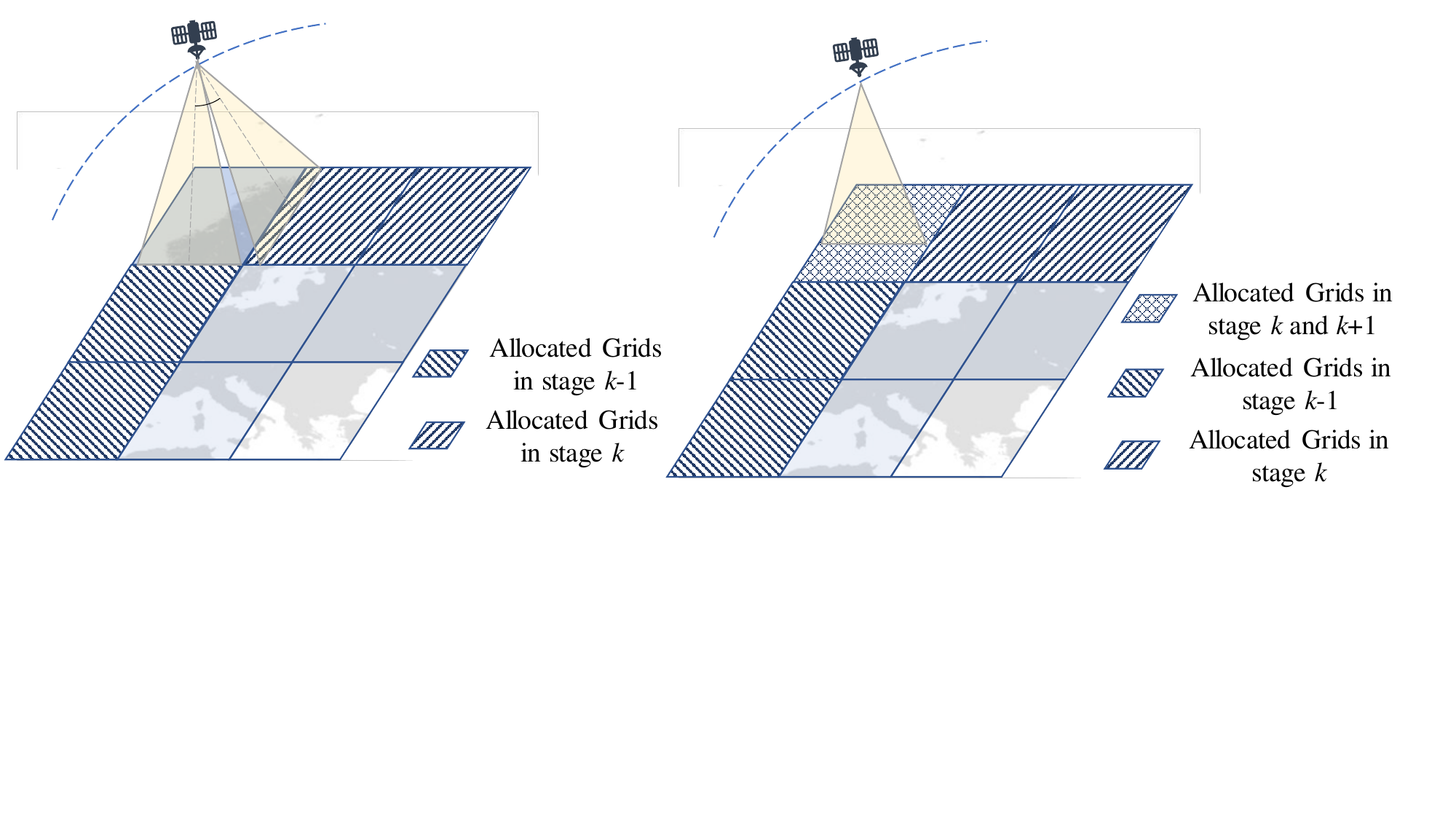}
    \label{F3A}
    }
    \quad
    \subfigure[Transition time ${{\eta }_{ik}}=0$]{
        \includegraphics[scale=0.48]{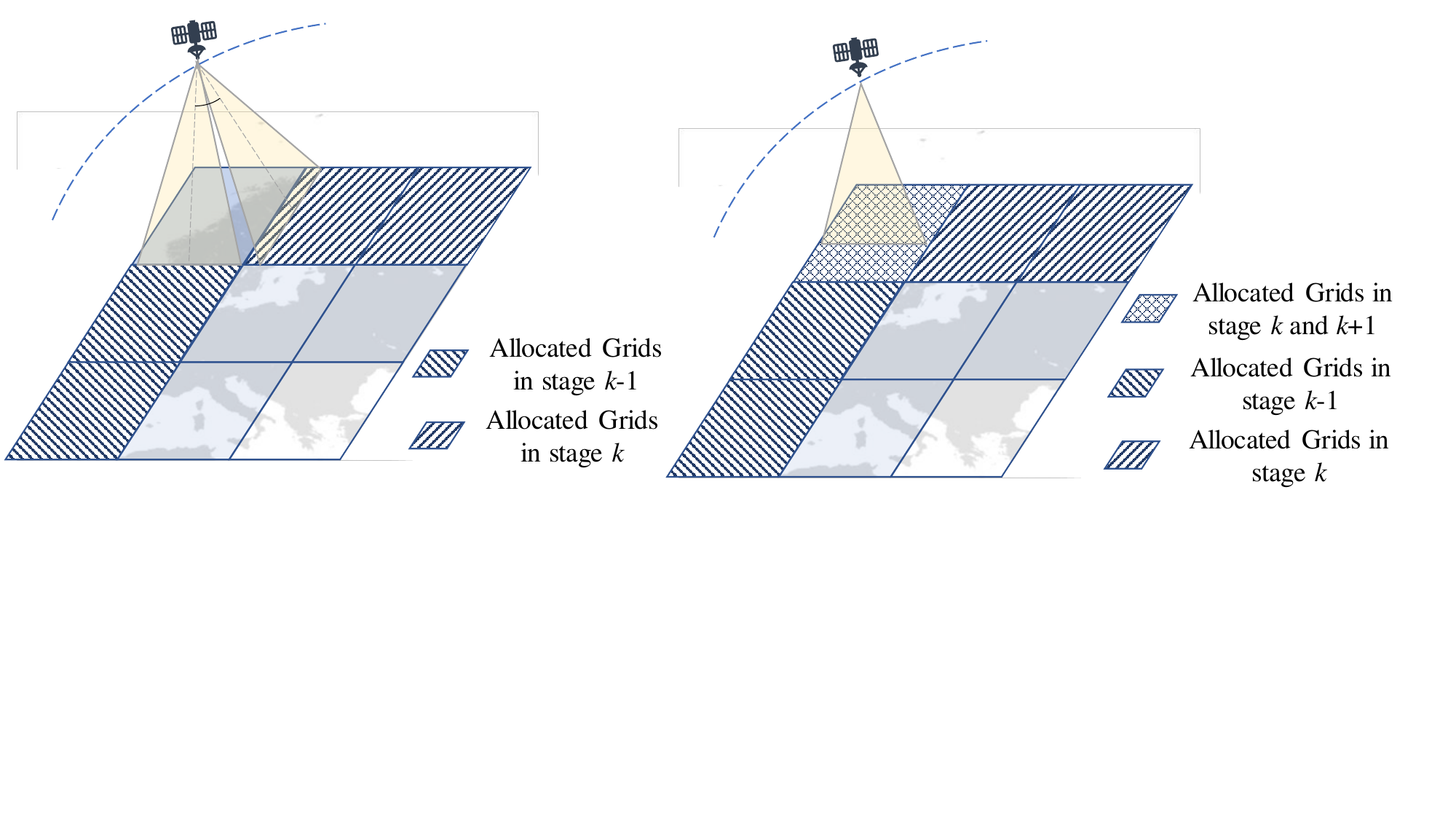}
    \label{F3B}
    }
\caption{The diagram of transition time}
\label{f3}
\end{figure}

Suppose $R_{ik}^{*}$ refers to the set of allocated grids of satellite $s_i$ in stage $k$ ($R_{ik}^{*}\subseteq {{R}_{ik}}$). If $R_{i(k-1)}^{*}\cap R_{ik}^{*}=\varnothing $, satellite $s_i$   takes some time to transfer the angle of the payload from the grid in $R_{i(k-1)}^{*}$ to the grid in $R_{ik}^{*}$. Thus, the payload transition time ${{\eta }_{ik}}$ for satellite $s_i$ in stage $k$ as defined as:
\begin{equation}
\begin{array}{l}
{{\eta }_{ik}}=\left\{ \begin{matrix}
   H\quad if\ R_{i(k-1)}^{*}\cap R_{ik}^{*}=\varnothing   \\
   0\ \quad if\ R_{i(k-1)}^{*}\cap R_{ik}^{*}\ne \varnothing   \\
\end{matrix} \right.
\end{array}
\label{eq2}  
\end{equation}
where $H$ is a positive constant,which is an approximation value. It simplifies the complex dynamics of real-world scenarios where transition times might indeed vary for different grids. As shown in the Fig. \ref{f3}, the transition time over two stages is $H$ when there is no intersection between the allocated grids of satellite $s_i$ in the stage $k-1$ and stage $k$, otherwise, it is 0.

By partitioning the timeline of the DGAP, the formulation described above decomposes it into sDGAPs to address its dynamic changes. The relation between adjacent sDGAP is specifically manifested as transition time between them, which means generating allocation files for the next stage depends on the allocation solution from the previous stage. Therefore, sequentially resolving the sDGAPs can effectively address the DGAP.

\subsubsection{Formulation of sDGAP}
For simplicity of notation, we omit the stage index ‘$k$’ until the multiple stage situation is considered in Section 4.3. In the sDGAP, $R_i$ represents the set of visible grids for satellite $s_i$, and $S_j$  represents the set of visible satellites for grid $r_j$. Each $r_j$  requires a satellite to complete the observation load $\beta_i$. The observation capacity of satellite $s_i$ toward grid $r_j$ is denoted $\alpha_{ij}$, and the time that satellite $s_i$ allocates to the grid $r_j$ is denoted $x_{ij}$. Let $y_j$ to be the remain observation load of grid $r_j$. We thus have
\begin{equation}
{{y}_{j}}={{\beta }_{j}}-\sum\limits_{j\in S_{j}^{*}}{{{\alpha }_{ij}}\cdot {{x}_{ij}}}
\label{eq3}  
\end{equation}
Additionally, if a satellite is allocated to more than two grids, then the imaging transition time of the satellite between grids needs to be considered. Let $R_{i}^{*}$ represents the allocated grid set of satellite $s_i$ ($R_{ik}^{*}\subseteq {{R}_{ik}}$). For satellite $s_i$, its imaging transition time  $\rho_i$ is calculated as
\begin{equation}
{{\rho }_{i}}=\left| R_{i}^{*} \right|\cdot C
\label{eq4}  
\end{equation}
where $|R_{i}^{*}|$ represent the cardinality of set $R_i^*$,  which is the number of grids to which satellite $s_i$ is allocated. $C$ is a constant that represents the unit conversion time.

Overall, we set the optimization objective of the problem to minimize the maximum number of remaining observations. The satellite observation capacity is different for each grid, and thus to accomplish the observation load of each grid, the allocation of satellites and grids requires consideration of the satellite observation capacity and grid observation load. Therefore, according to the above analysis, in stage $k$, the problem model P0 is established as 
\begin{equation}
\min \ \underset{j\in M}{\mathop{\max }}\,\left( {{\beta }_{j}}-\alpha _{j}^{\text{T}}x \right)
\label{eq5}  
\end{equation}
s.t. 
\begin{equation}
\sum\limits_{j\in R_{i}^{*}}{{{x}_{ij}}}+{{\rho }_{i}}\le \Delta t\quad \forall {{s}_{i}}\in S
\label{eq6}  
\end{equation}

where $x_{ij}$ is a decision variable representing the time that satellite $s_i$ is allocated to grid $r_j$ and $\Delta t$ is the total allocatable time in this stage. Formulated as Eq. \ref{eq5}, the objective function is set to minimize the maximum number of remaining observations, where $\alpha _{j}^{T}x=\sum\limits_{j\in S_{j}^{*}}{{{\alpha }_{ij}}}\cdot {{x}_{ij}}$ represents the observation load for gird $j$ in the allocation file $x$. Formulated as Eq. \ref{eq6}, the constraint states that the sum of the transition time $\rho_i$ and total allocation time of satellite $s_i$ are not greater than the allocatable time $\Delta t$.

\subsection{Global utility function}
Considering the above problem P0, the global utility function is defined as follows:
\begin{equation}
U=\sum\limits_{j\in M}{{{-e}^{\frac{1}{\varepsilon }({{\beta }_{j}}-\alpha _{j}^{\text{T}}x)}}}
\label{eq7}  
\end{equation}

\textbf{Claim 1.} Let $h(x)=\varepsilon \log \left( \sum\limits_{j\in M}{{{e}^{\frac{1}{\varepsilon }({{\beta }_{j}}-\alpha _{j}^{\text{T}}x)}}} \right)$. Suppose the problem P1 has the objective function \[\text{minimize}\quad h(x)\]
the optimal value of problem P1 is $p_{1}^{*}$ and the optimal value of problem P0 is $p_{0}^{*}$, we then have $0\le p_{1}^{*}-p_{0}^{*}\le \varepsilon \log m$.

\textbf{Proof}: We provide a complete proof for the Claim. 1 in the Appendix. Note that the parameter $\varepsilon$ is a positive constant that regulates the degree of approximation between the objective function of problem P0 and that of problem P1. When $\varepsilon$ approaches zero, the objective function of P0 can be approximate by minimizing $h(x)$.

\textbf{Theorem 1.} Suppose $U(x)={{-e}^{\frac{1}{\varepsilon }h(x)}}$=$\sum\limits_{j\in M}{{{-e}^{\frac{1}{\varepsilon }({{\beta }_{j}}-\alpha _{j}^{\text{T}}x)}}}$, the objective function of the original problem P0 can be approximated by $\text{maximize}\ U(x)$.

\textbf{Proof}: First, according to Claim 1, the objective function of P0 can be approximated by the objective function of P1. Suppose $x^*$ is the optimal solution for \[\text{maximize}\ U(x)\]
It is then required to prove that $x^*$ is also the optimal solution of P1.
We prove the above claim through contradiction. Suppose $x^*$ is not the optimal solution for problem P1, which means there exit $h(\bar{x})<h({{x}^{*}})$($\bar{x}\in \mathbb{R}$). ${{-e}^{{}^{1}\!\!\diagup\!\!{}_{\varepsilon }\;}}(\varepsilon >0)$ is a monotone decreasing function, and thus, if $h(\bar{x})<h({{x}^{*}})$, then $-{{e}^{\frac{1}{\varepsilon }h(\overline{x})}}>-{{e}^{\frac{1}{\varepsilon }h(x*)}}$. This means that there exists $U(\bar{x})>U({{x}^{*}})$ ($\bar{x}\in \mathbb{R}$), which contradicts the condition that $x^*$ is the optimal solution. And this completes the proof.

\subsection{Potential game model}
Let $G=\left\{ N,\left\{ {{A}_{i}} \right\},\left\{ {{U}_{i}} \right\} \right\}$ denote a game, where $N=\left\{ 1,2,\ldots ,n \right\}$ is a set of $n$ agents, $A={{A}_{1}}\times {{A}_{2}}\ldots \times {{A}_{n}}$ is the joint action set, and ${{U}_{i}}:{{A}_{i}}\to R$ is the local utility function of agent $i$. For an action file $a=\left( {{a}_{1}},{{a}_{2}},\ldots ,{{a}_{n}} \right)\in A$, where ${{a}_{i}}\in {{A}_{i}}$ is an action of agent $i$, $a_i$ represents the allocation file of satellite $s_i$.

In the problem description, the non-continuous objective function in problem P0 is approximated by employing Eq. \ref{eq7} as a global utility function. To ensure the action file satisfies the constraints of P0, the actions in the action space has been initially selected. Specifically, when constructing the action file in the potential game model, it is necessary to filter out actions that comply with the constraints of P0, which is given by Eq. \ref{eq6}.

\begin{equation}
{{a}_{i}}=\left( {{x}_{i1}},{{x}_{i2}},\ldots ,{{x}_{ij}},\ldots ,{{x}_{im}} \right)
\label{eq8}  
\end{equation}
Here, $x_{ij}$ is the time that satellite $s_i$ allocate to grid $e_j$. Let ${{a}_{-i}}=\left( {{a}_{1}},{{a}_{2}}\ldots ,{{a}_{i-1}},{{a}_{i+1}},\ldots ,{{a}_{n}} \right)$ to be the profile of agent actions other than those of satellite $s_i$, and $a_{i}^{0}$ represents a specific null action for each participant $i$. Consequently, every $x_ij$ within the $a_{i}^{0}$ equals to zero.

According to the global utility function $U(x)={{-e}^{\frac{1}{\varepsilon }h(x)}}$=$\sum\limits_{j\in M}{{{-e}^{\frac{1}{\varepsilon }({{\beta }_{j}}-\alpha _{j}^{\text{T}}x)}}}$, a local utility function for an individual base on the wonderful life Utility rule \cite{DavidH.Wolpert.1999} is designed as  
\begin{equation}
\begin{aligned}
  & U({{a}_{i}},{{a}_{-i}})-U(a_{i}^{0},{{a}_{-i}}) \\ 
 & =\sum\limits_{j\in {{N}_{{{a}_{i}}}}}{\left( {{e}^{\frac{1}{\varepsilon }\left( {{\beta }_{j}}-\sum\limits_{i'\in S_{a}^{j}/\{i\}}{{{x}_{i'j}}{{\alpha }_{i'j}}} \right)}}-{{e}^{\frac{1}{\varepsilon }\left( {{\beta }_{j}}-\sum\limits_{i'\in S_{a}^{j}}{{{x}_{i'j}}{{\alpha }_{i'j}}} \right)}} \right)} \\ 
 & =\sum\limits_{j\in {{N}_{{{a}_{i}}}}}{\left( \left( {{e}^{-\frac{1}{\varepsilon }\sum\limits_{i'\in S_{a}^{j}/\{i\}}{{{x}_{i'j}}{{\alpha }_{i'j}}}}}-{{e}^{-\frac{1}{\varepsilon }\sum\limits_{i'\in S_{a}^{j}}{{{x}_{i'j}}{{\alpha }_{i'j}}}}} \right)\cdot {{e}^{\frac{1}{\varepsilon }{{\beta }_{j}}}} \right)} \\ 
\end{aligned}
\label{eq9}  
\end{equation}
where ${{N}_{{{a}_{i}}}}=\left\{ {{r}_{j}}\text{ }\!\!|\!\!\text{ }{{x}_{ij}}>0,{{x}_{ij}}\in {{a}_{i}} \right\}$ is the set of grids that $s_i$ allocates in the action $a_i$, and $S_{a}^{j}=\left\{ {{s}_{i}}\text{ }\!\!|\!\!\text{ }{{x}_{ij}}>0,{{x}_{ij}}\in {{a}_{i}},\forall {{a}_{i}}\in a \right\}$ is the satellite set that allocated to the grid $r_j$ in an action file $a$.

Then, to normalize the utility function
 \begin{equation}
\begin{array}{c}
\sum\limits_{j \in {N_{{a_i}}}} {\left( {\left( {{e^{ - \frac{1}{\varepsilon }\sum\limits_{i' \in S_a^j/\{ i\} } {{x_{i'j}}{\alpha _{i'j}}} }} - {e^{ - \frac{1}{\varepsilon }\sum\limits_{i' \in S_a^j} {{x_{i'j}}{\alpha _{i'j}}} }}} \right) \cdot {e^{\frac{1}{\varepsilon }{\beta _j}}}} \right)} \\
 = \sum\limits_{j \in {N_{{a_i}}}} {\left( {\left( {1 - {e^{ - \frac{1}{\varepsilon }{x_{ij}}{\alpha _{ij}}}}} \right){e^{ - \frac{1}{\varepsilon }\sum\limits_{i' \in S_a^j/\{ i\} } {{x_{i'j}}{\alpha _{i'j}}} }} \cdot {e^{\frac{1}{\varepsilon }{\beta _j}}}} \right)} \\
 = \sum\limits_{j \in {N_{{a_i}}}} {\left( {\left( {1 - {e^{ - \frac{1}{\varepsilon }{x_{ij}}{\alpha _{ij}}}}} \right) \cdot {e^{\frac{1}{\varepsilon }\left( {{\beta _j} - \sum\limits_{i' \in S_a^j/\{ i\} } {{x_{i'j}}{\alpha _{i'j}}} } \right)}}} \right)} \\
 \le {N_{\max }} \cdot \left( {1 - {e^{ - \frac{1}{\varepsilon } \cdot t{\alpha _{\max }}}}} \right){e^{\frac{1}{\varepsilon } \cdot {\beta _{\max }}}}
\end{array}
 \end{equation}

Here, ${{\alpha }_{max}}$ is the maximal satellite capacity, ${{\beta }_{max}}$ is the maximal observation load of the grid, and ${N_{\max}}$ is the maximal number of the grids that can be allocated:
\begin{equation}
{{\alpha }_{max}}=\underset{i,j}{\mathop{\max }}\,{{\alpha }_{ij}},{{\beta }_{max}}=\underset{j}{\mathop{\max }}\,{{\beta }_{j}},{N_{\max }} = \mathop {\max }\limits_i {\mkern 1mu} \left| {{N_{{a_i}}}} \right|
\label{eq11}  
\end{equation}
Letting $P = {N_{\max }}\left( {1 - {e^{ - \frac{{t{\alpha _{\max }}}}{\varepsilon }}}} \right){e^{\frac{{{\beta _{\max }}}}{\varepsilon }}}$, we can obtain the local utility function of satellite $s_i$ as
\begin{equation}
{{U}_{i}}\left( {{a}_{i}},{{a}_{-i}} \right)=\frac{1}{P}\underset{j\in {{N}_{{{a}_{i}}}}}{\mathop \sum }\,\left( \left( {{e}^{-\frac{1}{\varepsilon }\underset{i\text{ }\!\!'\!\!\text{ }\in S_{a}^{j}/\left\{ i \right\}}{\mathop \sum }\,{{x}_{i\text{ }\!\!'\!\!\text{ }j}}{{\alpha }_{i\text{ }\!\!'\!\!\text{ }j}}}}-{{e}^{-\frac{1}{\varepsilon }\underset{i\text{ }\!\!'\!\!\text{ }\in S_{a}^{j}}{\mathop \sum }\,{{x}_{i\text{ }\!\!'\!\!\text{ }j}}{{\alpha }_{i\text{ }\!\!'\!\!\text{ }j}}}} \right)\cdot {{e}^{\frac{1}{\varepsilon }{{\beta }_{j}}}} \right)
\label{eq12}  
\end{equation}
The Nash equilibrium is a core solution of non-cooperative games and represents situations in which no player has an incentive to unilaterally deviate from the game rules. The Nash equilibrium is formally defined as follows.

\textbf{Definition 1} For a game $G=\left\{ N,\left\{ {{A}_{i}} \right\},\left\{ {{U}_{i}} \right\} \right\}$, an action ${{a}^{N}}$ is call a Nash equilibrium if for each agent $i$,
\begin{equation}
{{U}_{i}}\left( a_{i}^{N},a_{-i}^{N} \right)=\underset{{{a}_{i}}\in {{A}_{i}}}{\mathop{\max }}\,{{U}_{i}}\left( {{a}_{i}},a_{-i}^{N} \right)
\label{eq13}
\end{equation}

\textbf{Definition 2} A game $G=\{ N,\left\{ {{A}_{i}} \right\}$ is an exact potential game if there is a function $\phi :{{A}_{i}}\to R$ satisfying
\begin{equation}
\begin{aligned}
  & {{U}_{i}}\left( {{{{a}'}}_{i}},{{a}_{-i}} \right)-{{U}_{i}}\left( {{{{a}''}}_{i}},{{a}_{-i}} \right)=\phi \left( {{{{a}'}}_{i}},{{a}_{-i}} \right)-\phi \left( {{{{a}''}}_{i}},{{a}_{-i}} \right) \\ 
 & \forall i\in N,\forall {{a}_{-i}}\in {{A}_{-i}},\forall {{{{a}'}}_{i}},{a}''\in {{A}_{i}} \\ 
\end{aligned}
\label{eq14}
\end{equation}

\textbf{Theorem 2.} The task allocation game $G=\{ N,\left\{ {{A}_{i}} \right\}$ is a potential game if the potential function $\phi $ are defined as
\begin{equation}
\phi =\frac{1}{P}U=\frac{1}{P}\underset{j=1}{\overset{m}{\mathop \sum }}\,{{e}^{\frac{1}{\varepsilon }\left( {{\beta }_{j}}-\alpha _{j}^{T}x \right)}}
\label{eq15}
\end{equation}

\textbf{Proof}: According to the definition of local utility function, we have:
\begin{equation}
{{U}_{i}}\left( {{a}_{i}},{{a}_{-i}} \right)=\frac{1}{P}{{U}_{i}}\left( {{a}_{i}},{{a}_{-i}} \right)-\frac{1}{P}{{U}_{i}}\left( a_{i}^{0},{{a}_{-i}} \right)
\label{eq16}
\end{equation}
\begin{equation}
{{U}_{i}}\left( a_{i}^{\text{ }\!\!'\!\!\text{ }},{{a}_{-i}} \right)=\frac{1}{P}{{U}_{i}}\left( a_{i}^{\text{ }\!\!'\!\!\text{ }},{{a}_{-i}} \right)-\frac{1}{P}{{U}_{i}}\left( a_{i}^{0},{{a}_{-i}} \right)
\label{eq17}
\end{equation}
Then,
\begin{equation}
\begin{aligned}
  & {{U}_{i}}\left( {{a}_{i}},{{a}_{-i}} \right)-{{U}_{i}}\left( a_{i}^{\text{ }\!\!'\!\!\text{ }},{{a}_{-i}} \right) \\ 
 & =\frac{1}{P}\left( U\left( {{a}_{i}},{{a}_{-i}} \right)-U\left( a_{i}^{0},{{a}_{-i}} \right) \right)-\frac{1}{P}\left( U\left( a_{i}^{\text{ }\!\!'\!\!\text{ }},{{a}_{-i}} \right)-U\left( a_{i}^{0},{{a}_{-i}} \right) \right) \\ 
 & =\frac{1}{P}\left( U\left( {{a}_{i}},{{a}_{-i}} \right)-U\left( a_{i}^{0},{{a}_{-i}} \right)+U\left( a_{i}^{0},{{a}_{-i}} \right)-U\left( a_{i}^{\text{ }\!\!'\!\!\text{ }},{{a}_{-i}} \right) \right) \\ 
 & =\frac{1}{P}U\left( {{a}_{i}},{{a}_{-i}} \right)-\frac{1}{P}U\left( a_{i}^{\text{ }\!\!'\!\!\text{ }},{{a}_{-i}} \right)=\phi \left( {{a}_{i}},{{a}_{-i}} \right)-\phi \left( a_{i}^{\text{ }\!\!'\!\!\text{ }},{{a}_{-i}} \right) \\ 
\end{aligned}
\label{eq18}
\end{equation}
According to the Definition 2, $\phi =\frac{1}{P}U$ is the potential function for game $G$.

\textbf{Theorem 3.} For the task allocation game $G=\{ N,\left\{ {{A}_{i}} \right\}$, the optimal solution $a^*$ of the problem P0 is the Nash equilibrium $a_N$ of the potential game $G$.

\textbf{Proof}: We prove the above theorem through contradiction: Suppose the optimal solution ${{a}^{*}}=\left( a_{1}^{*},a_{2}^{*},\ldots ,a_{n}^{*} \right)$ is not the Nash equilibrium of the potential game $G=\{ N,\left\{ {{A}_{i}} \right\}$, which means that there exists for which
\begin{equation}
{{U}_{i}}\left( a_{i}^{\text{ }\!\!'\!\!\text{ }},{{a}_{-i}} \right)-{{U}_{i}}\left( a_{i}^{\text{*}},{{a}_{-i}} \right)>0
\label{eq19}
\end{equation}
According to Definition 2, we have
\begin{equation}
\begin{aligned}
  & {{U}_{i}}\left( {{{{a}'}}_{i}},{{a}_{-i}} \right)-{{U}_{i}}\left( {{{{a}''}}_{i}},{{a}_{-i}} \right)=\phi \left( {{{{a}'}}_{i}},{{a}_{-i}} \right)-\phi \left( {{{{a}''}}_{i}},{{a}_{-i}} \right)\  \\ 
 & \forall i\in N,\forall {{a}_{-i}}\in {{A}_{-i}},\forall {{{{a}'}}_{i}},{a}''\in {{A}_{i}} \\ 
\end{aligned}
\label{eq20}
\end{equation}
Thus,
\begin{equation}
\begin{aligned}
  & {{U}_{i}}\left( {{{{a}'}}_{i}},{{a}_{-i}} \right)-{{U}_{i}}\left( a_{i}^{\text{*}},{{a}_{-i}} \right) \\ 
 & =\phi \left( {{{{a}'}}_{i}},{{a}_{-i}} \right)-\phi \left( a_{i}^{\text{*}},{{a}_{-i}} \right) \\ 
 & =\frac{1}{P}\left( U\left( {{{{a}'}}_{i}},{{a}_{-i}} \right)-U\left( a_{i}^{\text{*}},{{a}_{-i}} \right) \right)>0 \\ 
\end{aligned}
\label{eq21}
\end{equation}
Therefore, $U\left( {{{{a}'}}_{i}},{{a}_{-i}} \right)-U\left( a_{i}^{*},{{a}_{-i}} \right)>0$, which contradicts to the condition that $a^*$ is the optimal solution of the problem P0. And this completes the proof.

\section{Algorithm design}
Numerous learning algorithms, such as fictitious play \cite{Shamma.2005}, best response \cite{Ai.2008}, log-linear learning \cite{Sun.2018} and other \cite{Shahid.2014} algorithms, exhibit good convergence. Among them, the better reply process (BRP) algorithm, introduced by Young \cite{Young.2004}, has been demonstrated to converge to equilibrium in potential games. However, the fixed parameter restricts the performance of the BRP algorithm because it fails to balance exploration and exploitation. Additionally, selecting the action with the better regret value requires the calculation of the efficiency function of all actions in each iteration, resulting in a high computational cost. To improve the performance of the BRP algorithm, a selective time-variant better reply process (SeTVBRP) algorithm is designed for the sDGAP. The SeTVBRP algorithm is proved to converge to equilibrium as time goes to infinity.

\subsection{Selective Time Variant Better Reply Process}
To solve the sDGAP, we first initialize the allocation file for all satellites following the greedy policy. The main coordination process is then implemented in a relay manner. Specifically, satellite ${{s}_{i-1}}$ receives allocation file $a(t-1)$ from satellite ${{s}_{i-1}}$ at iteration $t$. After updating the time-variant parameters $\varepsilon(t)$ and $\omega \left( t \right)$, satellite ${{s}_{i-1}}$ selects an action having a utility function value better than that of the last action. A trial action is chosen from these better actions. Satellite ${{s}_{i-1}}$ updates action $a_i(t)$ by the trial action with probability $\vartheta $.

\begin{algorithm}[t]  
	\caption{Selective time variant better reply process}
	\LinesNumbered 
	\KwIn{Satellite set $S$ and action sat $A$}
	\KwOut{Final action file $a({T}_{max})$}
	  Initialize action file $a^0$ by greedy policy\; 
    \For{each iteration t=1,2,…,${{T}_{max}}$}{
        Select satellite ${{s}_{i}}\in S$\; 
        Update parameter $\varepsilon \left( t \right)$ and $\omega \left( t \right)$\; 
        Select part of action set $A_{i}^{t} \subseteq  A_i$ by ratio $\omega \left( t \right)$\;
        Exchange $a_{i}^{t-1}$ information with neighbors and calculate $U_{i}^{t-1}$\; 
        \For{action $j \in A_{i}^{t}$}{
            Calculate ${{U}_{i}}\left( a_{i}^{j},{{a}_{-i}} \right)$\;
            \If{${{U}_{i}}\left( a_{i}^{j},{{a}_{-i}} \right)>U_{i}^{t-1}$}{
                Add $a_{i}^{j}$ into the better action set $B_{i}^{t}$\;
            }
            
        }
        \eIf{$B_{i}^{t}\ne \varnothing $}{
            Choose a trial action ${{\hat{a}}_{i}}$ from $B_{i}^{t}$ randomly\;
            Update $a_{i}^{t}={{\hat{a}}_{i}}$ with probability $1-\vartheta $\;
        }
        {
            $a_{i}(t)=a_{i}(t-1)$\;
        }
    }
\end{algorithm}

In this process, the time-variant parameter method, selective action method and inertia method are introduced to improve the original BRP method.

The time-variant parameter approach draws inspiration from simulated annealing \cite{Wu.2017}, where the temperature $\zeta $ determines the amplitude of noise, indicating the likelihood of an agent taking a suboptimal action. As $\zeta $ approaches zero, an agent selects the best response action with high probability. We observe a similar effect on the parameter $\varepsilon $ with the utility function $U_i$, where a large $\varepsilon $ provides a global view of the utility function and the agent is inclined to take actions that reduce the overall observation load. However, as $\varepsilon $ approaches zero, the utility function $U_i$ is such that the agent is more inclined to select an action that directly optimizes the global goal. The time-variant parameter method ensures that $\varepsilon $ monotonically decreases with each iteration. This method ensures that the algorithm focuses on exploring the possibility of a better solution in early iterations and later exploits and improves the existing solution.

In the original BRP algorithm, the utility function of all actions needs to be calculated in each iteration, but only one action is selected in the end, which is a great waste of computing resources. Therefore, a selective action method is designed. The partial action set $A_{i}^{t}$ is extracted from the whole action set $A_i$ using the ratio $\omega \left( t \right)$ for calculation at each iteration, as indicated in line 5 of the Algorithm 1. Moreover, $\omega \left( t \right)$ gradually increases with the number of iterations. Here, $\omega \left( t \right)$ represents the proportion of filtered partial actions in $A_{i}^{t}$ relative to the total count of actions in $A_i$, and it increases with the number of iterations.

Lastly, we introduce a small amount of inertia into the learning process. In particular, a player takes the same action as in the previous period with probability $\vartheta $, and chooses a better reply action randomly from the selective actions set with probability of $1-\vartheta $.

\subsection{Convergence analysis}
The convergence analysis of the traditional BRP algorithm has been proved \cite{Young.2004} by constructing a weakly acyclic game under a better reply graph. This paper introduces the SeTVBRP algorithm, where the parameters $\varepsilon$ and $\omega$ are designed to be time variant. A new convergence analysis for the proposed SeTVBRP algorithm is thus required. In this section, we prove that our SeTVBRP algorithm converges to a Nash equilibrium surely for nearly any potential game.

First, two assumptions are made and a series of claims are stated and proved.

\textbf{Assumption 1.}There exist constants $\varepsilon_L$, $\varepsilon_U$, $\omega_L$ and $\omega_U$ such that
\begin{equation}
\varepsilon (t-1)\ge \varepsilon (t)
\label{eq22}
\end{equation}
\begin{equation}
\omega \left( t-1 \right)\le \omega \left( t \right)
\label{eq23}
\end{equation}
\begin{equation}
0\le {{\varepsilon }_{L}}\le \varepsilon (t)\le {{\varepsilon }_{U}}
\label{eq24}
\end{equation}
\begin{equation}
0<{{\omega }_{L}}\le \omega \left( t \right)\le {{\omega }_{U}}\le 1
\label{eq25}
\end{equation}
for all iteration $t>1$.

\textbf{Assumption 2.} The potential game defined in Theorem 3 only has only one Nash Equilibrium.\footnote{The hardware preconditions underpinning this assumption necessitate the real-time data exchange and decision-making, requiring collaboration among players based on the communication infrastructure and computational capabilities. And the theoretical condition for the assumption is that the potential function need to be strictly concave (or convex), which conforms to the potential function defined in Theorem 3.}

\textbf{Claim 2.} Consider a stochastic process with an infinite length. If event $\mathbb{E}$ happens with a positive probability ${{p}^{t}}\left( \mathbb{E} \right)$ at each time $t$, then $\mathbb{E}$ happens with probability 1.

\textbf{Proof.} Since the probability that $\mathbb{E}$ never happens is calculated by
\begin{equation}
{{p}_{0}}\left( \mathbb{E} \right)=\underset{T\to \infty }{\mathop{\lim }}\,\mathop{\prod }_{t=1}^{T}\left( 1-{{p}^{t}}\left( \mathbb{E} \right) \right)=0
\label{eq26}
\end{equation}
$\mathbb{E}$ happens with probability ${{p}_{1}}=1-{{p}_{0}}\left( \mathbb{E} \right)=1$.

\textbf{Claim 3.} Fix ${{t}_{0}}>{{T}_{f}},{{T}_{f}}=min\left\{ t:\varepsilon \left( t \right)={{\varepsilon }_{L}} \right\}$. For a potential game $G=\left\{ N,\left\{ {{A}_{i}} \right\},\left\{ {{U}_{i}} \right\} \right\}$, 
\begin{equation}
\phi \left( a(t) \right)-\phi \left( a(t_0) \right)\ge 0
\label{eq27}
\end{equation}
for all $t>t_0$.

\textbf{Proof.} Suppose the action agent $i$ selected at iteration ${{t}_{0}}+1$ is $a_{i}({{t}_{0}}+1)$. According to SeTVBRP, the next action always selects from the better action set or remain still, then we have 
\begin{equation}
\begin{aligned}
  & {{U}_{i}}({{a}_{i}}({{t}_{0}}+1),{{a}_{-i}}({{t}_{0}}+1))={{U}_{i}}({{a}_{i}}({{t}_{0}}+1),{{a}_{-i}}({{t}_{0}})) \\ 
 & \ge {{U}_{i}}({{a}_{i}}({{t}_{0}}),{{a}_{-i}}({{t}_{0}})) \\ 
\end{aligned}
\label{eq28}
\end{equation}
for each agent $i$. According to the Definition. 2, then
\begin{equation}
\begin{aligned}
  & \phi (a({{t}_{0}}+1))-\phi (a({{t}_{0}})) \\ 
 & ={{U}_{i}}({{a}_{i}}({{t}_{0}}+1),{{a}_{-i}}({{t}_{0}}+1))-{{U}_{i}}({{a}_{i}}({{t}_{0}}),{{a}_{-i}}({{t}_{0}}))\ge 0
\end{aligned}
\label{eq29}
\end{equation}
The argument can be repeated to show that $\phi (a(t))-\phi (a(t_0))\ge 0$ for all $t>{{t}_{0}}$.

\textbf{Claim 4.} Fix ${{t}_{0}}>{{T}_{f}}$. For a potential game $G=\left\{ N,\left\{ {{A}_{i}} \right\},\left\{ {{U}_{i}} \right\} \right\}$, assume $a(t_0)$ is Nash equilibrium. Then, $a(t)=a(t_0)$ for all $t>{{t}_{0}}$.

\textbf{Proof.} Since $a(t_0)$ is Nash equilibrium in a potential game $G$. According to Definition 1, we have
\begin{equation}
{{U}_{i}}\left( a_{i}(t_0),a_{-i}(t_0) \right)=\underset{{{a}_{i}}\in {{A}_{i}}}{\mathop{\max }}\,{{U}_{i}}\left( {{a}_{i}},a_{-i}(t_o) \right)
\label{eq30}
\end{equation}
for all agent $i$, where $A_i$ is the action set for agent $i$. Therefore, for any action ${{a}_{i}}\in {{A}_{i}}$ and any agent $i$,
\begin{equation}
{{U}_{i}}\left( {{a}_{i}},a_{-i}(t_o) \right)-{{U}_{i}}\left( a_{i}(t_0),a_{-i}(t_0) \right)\le 0
\label{eq31}
\end{equation}
That is, $\phi \left( a(t) \right)-\phi \left( a(t_0) \right)\le 0$ for all $t>{{t}_{0}}$. By Claim 3, we have $\phi \left( a(t) \right)-\phi \left( a(t_0) \right)\ge 0$ for all $t>{{t}_{0}}$. Thus,
\begin{equation}
\phi \left( a(t) \right)=\phi \left( a(t_0) \right)
\label{eq32}
\end{equation}
for all $t>{{t}_{0}}$ According to the Assumption 2, we have $a(t)=a(t_0)$ for all $t>{{t}_{0}}$.

\textbf{Claim 5.} Fix ${{t}_{0}}>{{T}_{f}}$. For a potential game $G=\left\{ N,\left\{ {{A}_{i}} \right\},\left\{ {{U}_{i}} \right\} \right\}$, assume $a(t_0)$ is not Nash equilibrium, and let $\hat{a}=({{\hat{a}}_{i}},a_{-i}(t_0))$ be such that ${{U}_{i}}\left( {\hat{a}} \right)>{{U}_{i}}\left( {{a}(t_0)} \right)$ for some agent $i$. $\hat{a}$ can be selected at iteration ${{t}_{0}}+N~\left( N\in {{\mathbb{R}}^{+}} \right)$ with at least probability $\gamma ={{\omega }_{L}}\cdot \frac{\left( 1-\vartheta  \right){{\vartheta }^{N-1}}}{\left| B_{i}(t_0) \right|}$.

\textbf{Proof.} In SeTVBRP algorithm, according to the selective action method and Assumption 1, ${{\hat{a}}_{i}}$ can be selected into $A_{i}^{t}$ for $t>{{t}_{0}}$ with at least probability ${{\omega }_{L}}$. Owing to ${{U}_{i}}\left( {\hat{a}} \right)>{{U}_{i}}\left( {{a}(t_0)} \right)$, agent $i$ select ${{\hat{a}}_{i}}$ for the next action with probability $\frac{1-\vartheta }{\left| B_{i}^{{{t}_{0}}} \right|}$, where $\left| B_{i}^{{{t}_{0}}} \right|$ is the number of actions that obtain a utility value better than ${{U}_{i}}\left( a(t_0) \right)$.

Owing to the inertia of agents, all other agent will repeat their action at iteration ${{t}_{0}}+N$ with probability ${{\vartheta }^{N-1}}$. This means the action file ${{\hat{a}}_{i}}$ can be selected at iteration ${{t}_{0}}+N~\left( N\in {{\mathbb{R}}^{+}} \right)$ with at least probability $\gamma ={{\omega }_{L}}\cdot \frac{\left( 1-\vartheta  \right){{\vartheta }^{N-1}}}{\left| B_{i}^{{{t}_{0}}} \right|}$.

\textbf{THEOREM 4.} In any potential game $G=\left\{ N,\left\{ {{A}_{i}} \right\},\left\{ {{U}_{i}} \right\} \right\}$ with Assumption 2, the SeTVBRP algorithm satisfying Assumption 1 converges in a finite number of iterations.

\textbf{Proof.} Fix ${{t}_{0}}>{{T}_{f}},{{T}_{f}}=min\left\{ t:\varepsilon \left( t \right)={{\varepsilon }_{L}} \right\}$. Suppose $a(t_0)$ is not Nash equilibrium. According to Claim 5, the better action file $a(t_1)$ will be reached at iteration ${{t}_{1}}:={{t}_{0}}+N$ with probability at least $\gamma $. Further suppose that $a(t_1)$ is not of Nash equilibrium, by Claim 5, the better action file $a(t_2)$ will be played at iteration ${{t}_{2}}:={{t}_{2}}+N$ with probability of at least $\gamma $.

The action set is finite, and we thus repeat the above argument till reaching the Nash equilibrium $a(t_L)$ at iteration $t_L$. From Claim 4, this would mean that the action file would stay at $a(t_L)$ for all $t>t_L$.

Therefore, given ${{t}_{0}}>{{T}_{f}}$, suppose ${{a}^{N}}$ is Nash equilibrium, there exist a positive probability and a positive integer $\bar{N}$, which are independent of ${{T}_{f}}$, for which the following event happens with at least probability $\bar{\gamma }:{{a}^{t}}={{a}^{N}}$ for all $t>{{t}_{0}}+\bar{N}$. Including Claim 2, we complete the proof that the SeTVBRP algorithm  satisfying Assumption 1 converges in a finite number of iterations with probability 1.

\subsection{Multi-stage dynamic allocation}
\begin{figure*}
	\centering
    \includegraphics[scale=0.35]{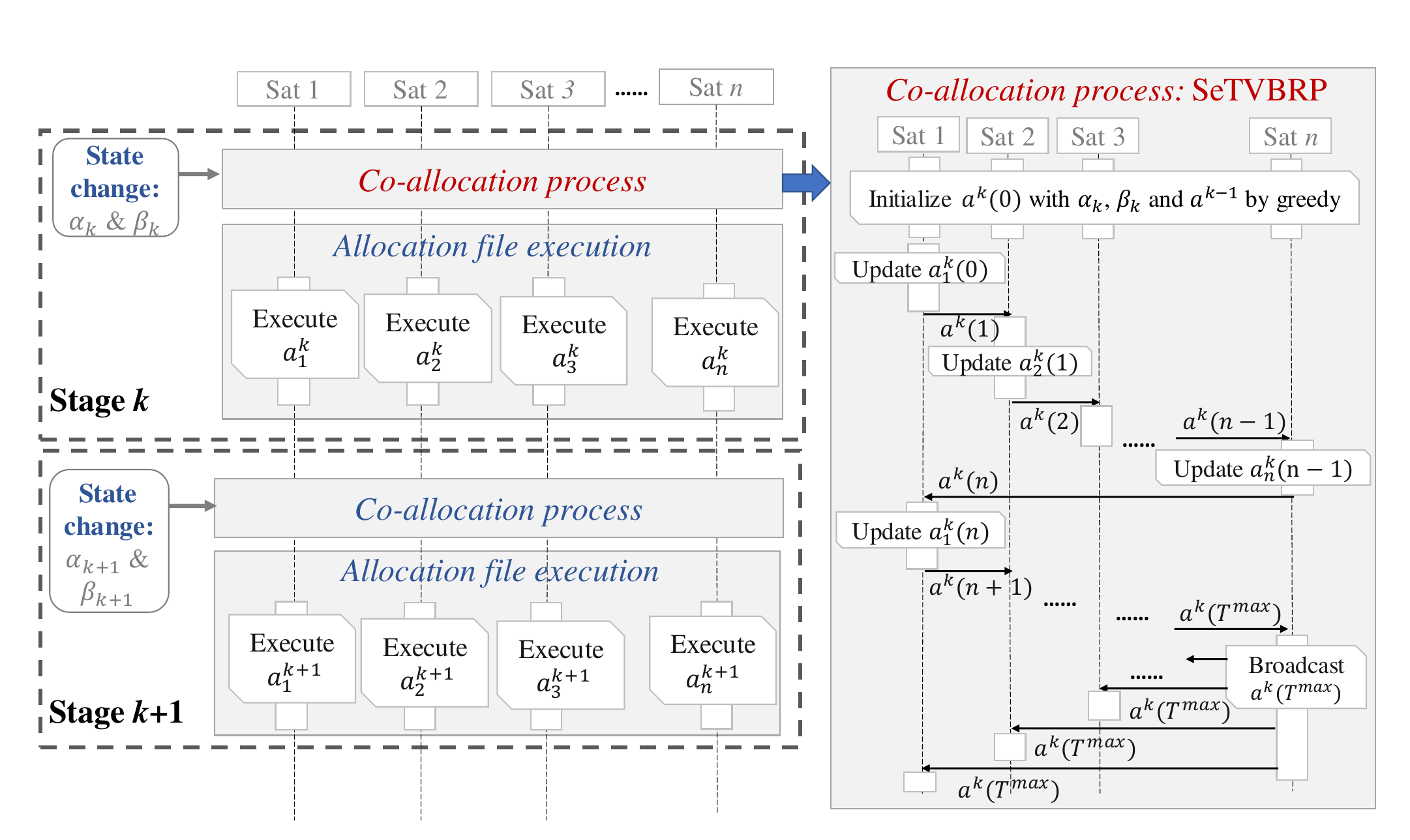}  
	\caption{Diagram for multi-stage dynamic allocation}   
\label{f4}
\end{figure*}

The previous section established an allocation model based on the potential game for the sDGAP of distributed satellites, whereas this section discusses the situation for multiple stages (i.e., the DGAP). As shown in Fig. \ref{f4}, a co\-allocation process is triggered when the state (i.e., the observation load $\beta$ or the observation capacity $\alpha$) changes. In stage $k$, the SeTVBRP algorithm is adopted to generate an allocation file $a^k$. The SeTVBRP algorithm firstly initializes an allocation file $a^k(0)$ with the new state parameters ($\alpha_k$ and $\beta_k$) and the allocation file $a^{k-1}$. The allocation file $a^{k-1}$ is used as input to compute the stage transition time $\eta_{ik} $ of satellite $s_i$. After the initial file $a^{k}(0)$ is generated, the satellite updates its allocation file and transmits it to the next satellite in a relay manner. Specifically, satellite $i$ transmits the updated allocation file to satellite $i'$, which then updates it and subsequently forwards it to satellite $i''$ until reaching satellite $n$. This communication relationship can be readily achieved within a satellite swarm, given that satellites typically equipped with at least two communication antennas are capable of establishing communication with adjacent satellites. After the updated allocation file $a^{k}(t)$ reaches Nash equilibrium or the maximum number of iterations $T^{max}$, the satellite that received the allocation scheme globally broadcasts the final allocation file $a^k$. This ends the co\-allocation process. The satellites then execute the final allocation file $a^k$ until another state change triggers the next stage.

\section{Simulation and comparison results}
In this section, experiments are first carried out for the SeTVBRP algorithm under different parameter settings, and the results of the SeTVBRP algorithm are compared to those of the original BRP algorithm to demonstrate the effectiveness of the improvement methods (i.e., selective action method and time-variant method). The superiority of the SeTVBRP algorithm is then further demonstrated by comparing the algorithm with state-of-the-art algorithms. Finally, the proposed algorithm is tested for the multiple-stage allocation situation in which the observation load increases with time. 

\subsection{Simulation settings}
To begin with, we consider a large-scale system having 150 satellites, which are distributed in five orbits in space. The right ascension of the ascending node changes from -24° to 24° in increments of 12°, where 30 satellites are distributed in each orbit. The difference in the argument of perigee between satellites in the same orbit is 12°. The satellites are connected by real-time inter-satellite links to achieve real-time communication. The basic parameters of the satellite system are given in Table 1.
\begin{table}[!ht]
    \centering
    \setlength{\belowcaptionskip}{0.2cm}
    \scriptsize
    \caption{The basic parameters of the satellite system}
    \begin{tabular}{ll}
    \hline
        Parameters & Value \\ \hline
        Semimajor axis  & 6878.14km  \\ 
        Eccentricity & 0  \\ 
        Number of orbital planes  &5  \\ 
        Number of satellites of each plan & 30  \\ 
        Orbit inclination angle of each plane  & 28.5°  \\ 
        The interval of the Right Asension of \\ 
        Ascending Node (RAAN) between each plane & 12°  \\
        Transition time constant $C$ & 1 \\
        Observation capacit $\alpha_{ik}$ & Random in [2, 3]  \\ \hline
    \end{tabular}
    \label{T1}
\end{table}

\begin{figure}[!htb]
    \centering
    \subfigure[Regional scenario]{
        \includegraphics[scale=0.28]{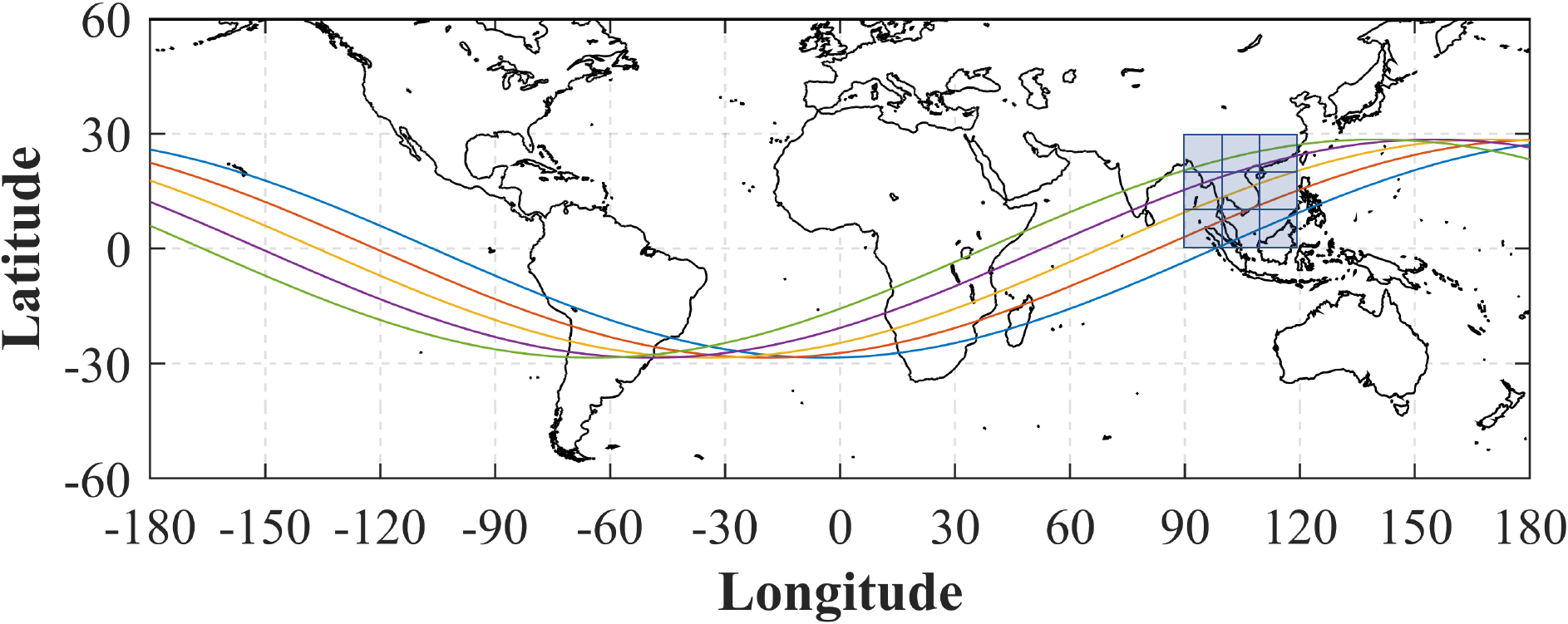}
    \label{F5A}
    }
    \quad
    \subfigure[Global scenario]{
        \includegraphics[scale=0.28]{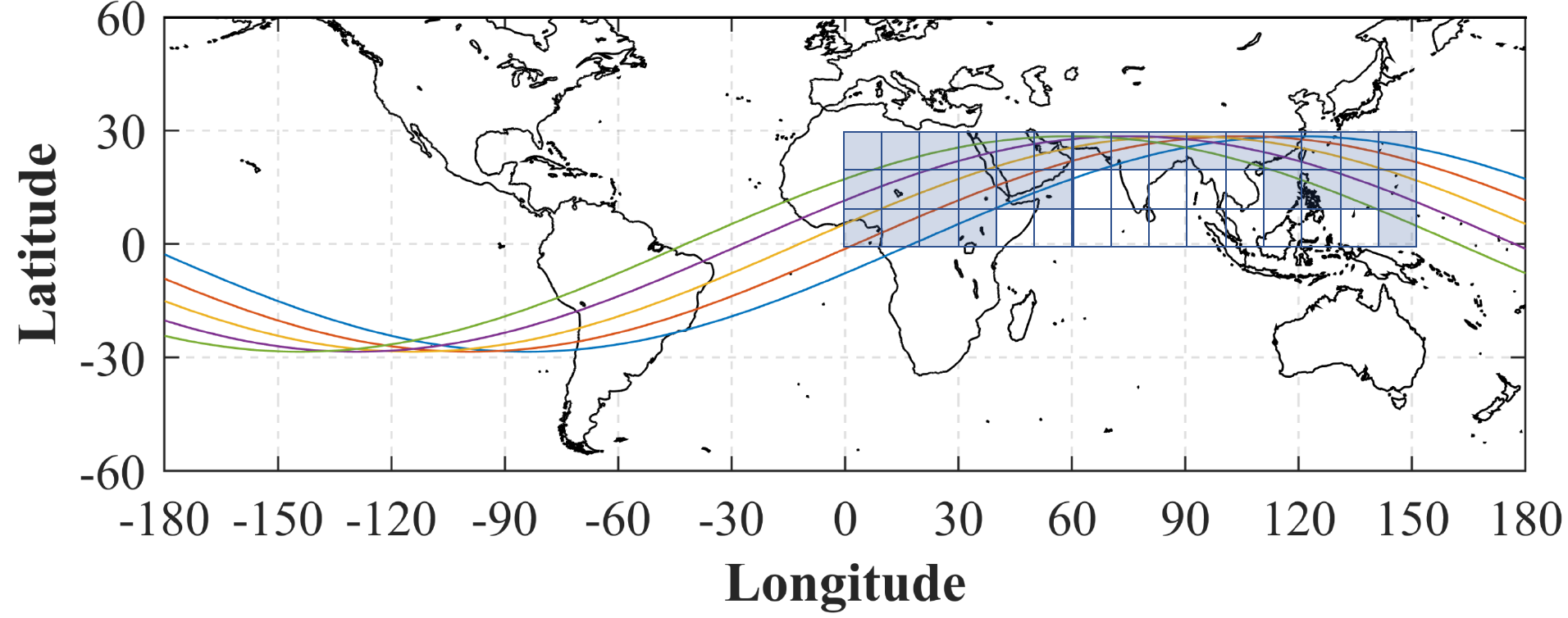}
    \label{F5B}
    }
\caption{Schematic diagram of scenario, where the grid represents the observation grid and the curve represents the ground track of the imaging satellite}
\label{f5}
\end{figure}

Two simulation scenarios are considered: (i) a regional scenario with nine grids and (ii) a global scenario with 30 grids. The System Tool Kit is used to obtain the visible time window among satellites and grids from 8:00 on 20 June 2022 to 9:00 on 20 June 2022. The observation load of the grids $\beta$ is randomly initialized as different positive values ranging from 30 to 80. The size of the grids depends on the satellite field of view, which is approximately 10° in latitude and longitude \cite{Wang.2020}. As shown in Fig \ref{f5} (a), the regional scenario is a square distribution of 3 × 3 grids with longitude 90°E–120°E and latitude 0°N–30°N. As shown in Fig. \ref{f5} (b), the global scenario is a set of 3 × 15 grids with longitude 90°E–120°W and latitude 0°N–30°N. Among these grids, only the 30 grids that lie below the satellite ground tracks are valid; i.e., (1) 3 × 4 grids with longitude 0°E–40°E and latitude 0°N–30°N, (2) 2 × 2 grids with longitude 40°E–60°E and latitude 10°N–30°N, (3) 1 × 5 grids with longitude 60°E–110°E and latitude 20°N–30°N, (4) 2 × 3 grids with longitude 110°E–140°E and latitude 10°N–30°N, and (5) 3 × 1 grids with longitude 140°E–150°E and latitude 0°N–30°N.

Note that only the satellites that are over a grid at the beginning of the simulation are involved in the allocation process. For the regional scenario, only 25 satellites participate in the allocation, whereas 100 satellites are involved for the global scenario.

\subsection{Parameter analysis experiments}
In this section, we discuss the performance of the SeTVBRP algorithm in solving the sDGAP. We set the decision interval $\text{ }\!\!\Delta\!\!\text{ }{{t}_{k}}=10$ min\footnote{In real-world scenarios, both capacity and load may fluctuate significantly during each $\text{ }\!\!\Delta\!\!\text{ }{{t}_{k}}$ duration. Thus, the condition of capacity and load constancy is an assumption based on a simplified model, and in case of considerable fluctuations within each $\text{ }\!\!\Delta\!\!\text{ }{{t}_{k}}$ duration, a smaller $\text{ }\!\!\Delta\!\!\text{ }{{t}_{k}}$ should be chosen to maintain the reliability of the results.}, which is also the available allocation time for a satellite. The minimum unit of time assigned in the experiment is minutes. We set the time-variant parameter $\varepsilon \left( t \right)$ and ratio $\omega \left( t \right)$ to be a piecewise linear nonincreasing function and nondecreasing function respectively
\begin{equation}
\varepsilon \left( t \right)=\left\{ \begin{matrix}
   {{\varepsilon }_{U}},\quad \quad \quad \quad \quad \quad \quad  \quad \quad\quad\quad \quad \quad t<\tau {{T}_{max}}  \\
   {{\varepsilon }_{U}}-\left( t-\tau {{T}_{max}} \right)\xi , \tau {{T}_{max}}\le t<\frac{{{\varepsilon }_{U}}-{{\varepsilon }_{L}}}{\varepsilon }+\tau {{T}_{max}}  \\
   {{\varepsilon }_{L}},\quad \quad \quad \quad \quad \quad \quad  \quad\quad\quad t\ge \frac{{{\varepsilon }_{U}}-{{\varepsilon }_{L}}}{\varepsilon }+\tau {{T}_{max}}  \\
\end{matrix} \right.
\label{eq33}
\end{equation}
\begin{equation}
\omega \left( t \right)=\left\{ \begin{matrix}
   t\varphi ,\quad \ \ t<\frac{1-\varphi }{\varphi }  \\
   1,\quad \quad t\ge \frac{1-\varphi }{\varphi }  \\
\end{matrix} \right.
\label{eq34}
\end{equation}
, where ${{\varepsilon }_{U}}$ and ${{\varepsilon }_{L}}$ are positive constant that represent the maximum and minimal value of ${{\varepsilon }(t)}$. And $t\varphi $ belongs to (0,1). ${{T}_{max}}$ refers to the maximum number of iterations.

We first investigate how the parameter $\tau $ affects the performance and convergence of the SeTVBRP algorithm. As shown in Eq. (22), the time-variant parameter  $\varepsilon \left( t \right)$ begins to decrease at iteration $\tau {{T}_{max}}$. The parameter $\tau $ directly affects the number of exploration and exploitation iterations. As the value of $\tau $ increases, the exploration phase in the initial iterations expands and the exploitation phase in the later iterations correspondingly shortens.

\begin{figure}
	\centering
    \includegraphics[scale=0.59]{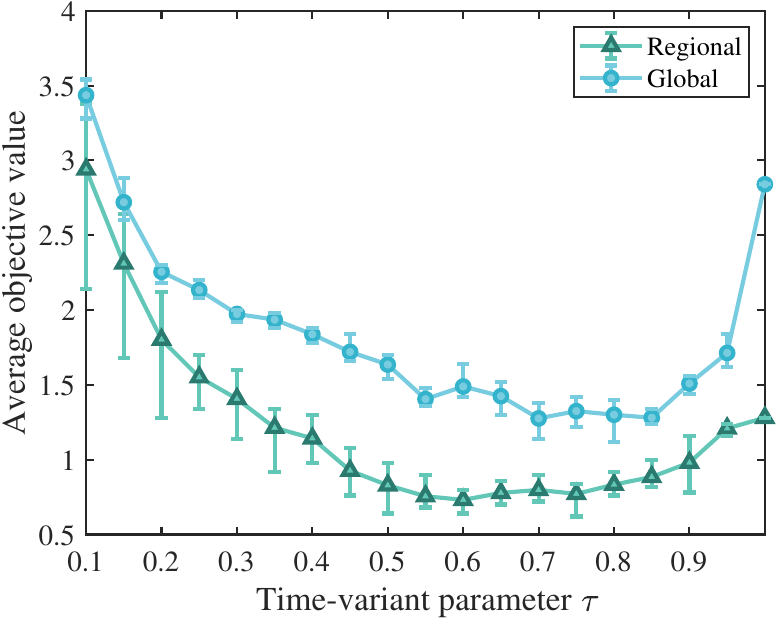}  
	\caption{ Allocation results for different values of the parameter $\tau $ compared to other algorithms}   
\label{f6}
\end{figure}

The remaining parameters are set as ${{\varepsilon }_{U}}$=15.4, ${{\varepsilon }_{L}}$=1, ${{\omega }_{L}}$=0.06,  ${{\omega }_{U}}$=1, $\varphi $=0.005. We set the time-variant parameter $\tau $ to change from 0 to 1 in increments of 0.05 and compare the solution objective values with error bars in both regional and global scenarios. Fig. \ref{f6} shows that the trend for the regional scenario is similar to that for the global scenario, with the objective value first decreasing and then increasing. The parameter range that provides better solutions is larger in the regional scenario (i.e., [0.55,0.9]) than in the global scenario (i.e., [0.7,0.85]).

\begin{figure*}[htbp]

    \centering
    \subfigure[Convergence curves]{
        \includegraphics[scale=0.59]{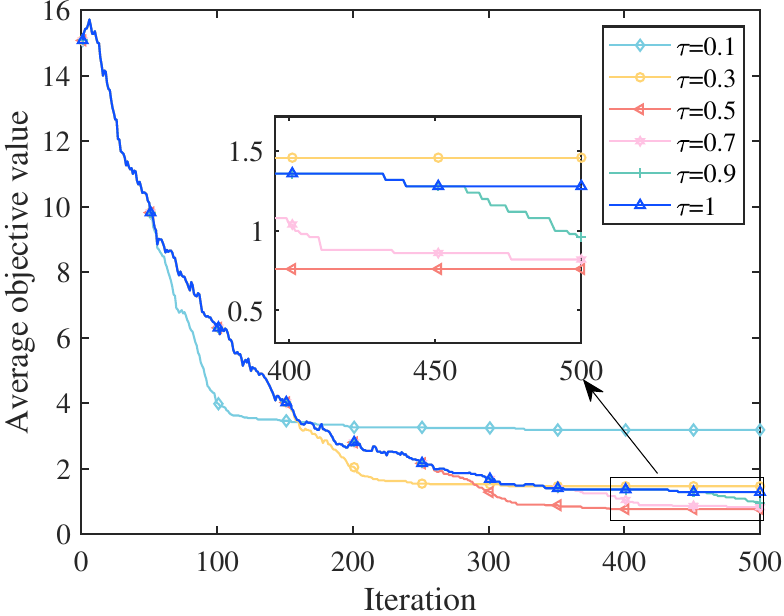}
    \label{f7a}
    }
    \quad
    \subfigure[Improvement in average objective value]{
        \includegraphics[scale=0.59]{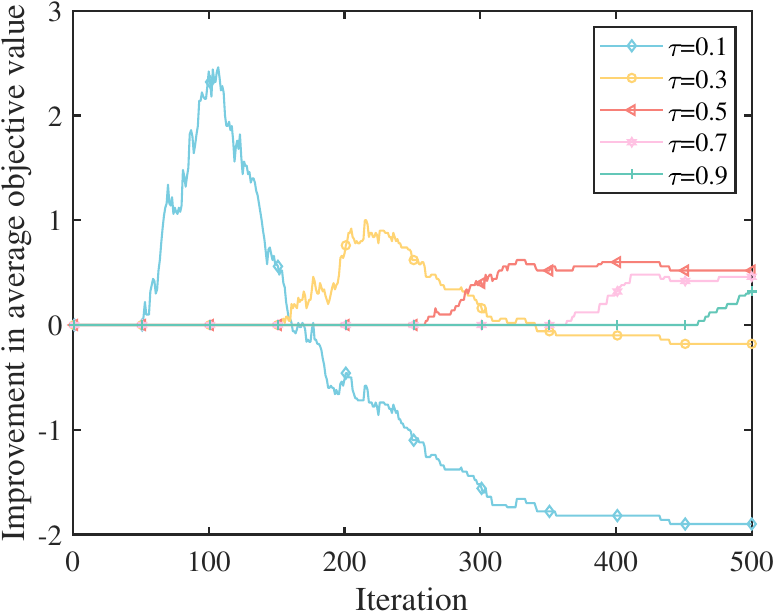}
    \label{f7b}
    }
\caption{Parameter analysis on the value of different time-variant parameters $\tau$ in the regional scenario}
\label{f7}
\end{figure*}

To further analyze the effect of the time-variant parameter $\tau $ on the convergence process, the convergence curve and the improvement curve of the convergence process are given for $\tau $ ranging from 0.1 to 0.9 in increments of 0.2. Fig. \ref{f7} show the convergence and improvement curves for the regional scenario, and Figs. 8 and 9 show those for the global scenario.

As shown in Fig. \ref{f7} (a), compared with the non-time-variant situation (i.e., $\tau $ = 1), when the number of iterations reaches $\tau {{T}_{max}}$, each curve enters the exploitation phase and convergence begins to accelerate; e.g., $\tau $ = 0.1 at $t$ = 50 and $\tau $ = 0.9 at $t$ = 450. Fig. \ref{f7} (b) presents the improvement in the objective value relative to the non-time-variant case to show this change more clearly. 

\begin{figure*}[htbp]

    \centering
    \subfigure[Convergence curves]{
        \includegraphics[scale=0.59]{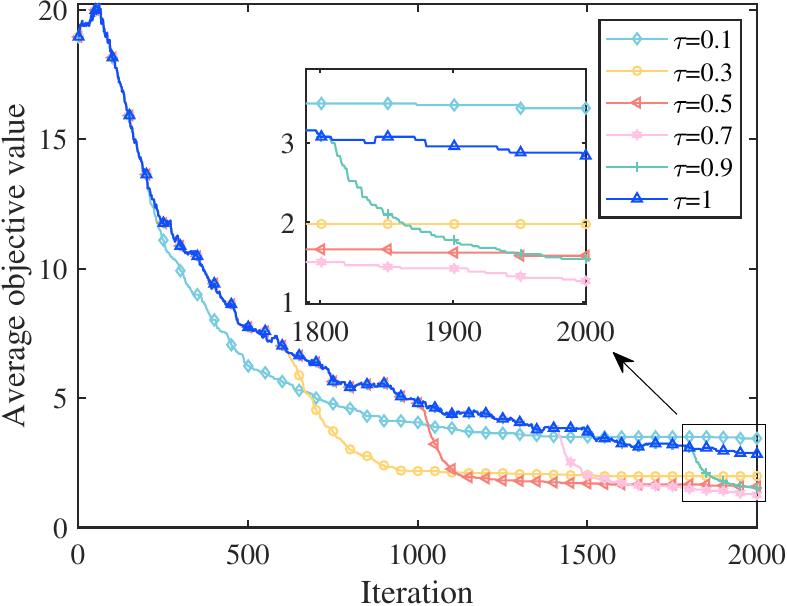}
    \label{f8a}
    }
    \quad
    \subfigure[Improvement in average objective value]{
        \includegraphics[scale=0.59]{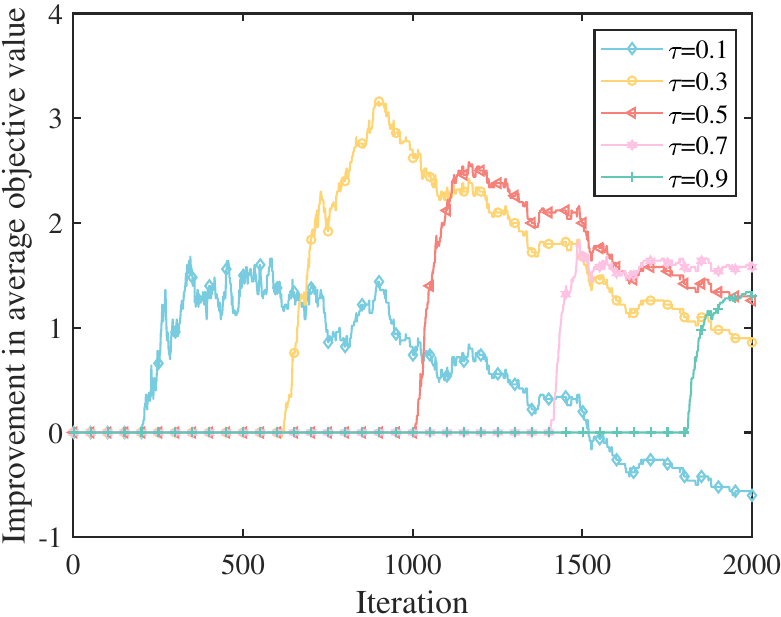}
    \label{f8b}
    }
\caption{Parameter analysis on the value of different time-variant parameters $\tau$ in the global scenario}
\label{f8}
\end{figure*}

The experimental results demonstrate that a diminutive $\tau $ value (such as 0.1 and 0.3) results in a relatively short exploration phase, limiting the lower bound of the solution even though there is a long solution exploitation phase in the subsequent stage. Conversely, an excessive $\tau $ value leads to an abbreviated exploitation stage, precluding adequate iterations to accomplish further optimization of the solution. Fig. \ref{f7} thus shows that the advantages of the time-variant method can be fully used by choosing an appropriate value of $\tau $ (approximately 0.5 to 0.7). The above conclusions on the time-variant parameter $\tau $ are more obvious in the global scenario, as shown in Fig. \ref{f8}. In the global scenario, the algorithm performs better when $\tau $ ranges from 0.6 to 0.8, which is a range slightly higher than that for the regional scenario. This result is mainly because increases in the numbers of grids and satellites increase the solution space, and a greater ratio of the exploration phase is thus required when searching for a better solution in the global scenario.

We also study how the time-variant speed $\xi$ affects the performance of the SeTVBRP algorithm. The results show that the time-variant speed $\xi$ affects the convergence curve only slightly in the early process and hardly affects the final convergence result. 

\begin{figure*}[htbp]

    \centering
    \subfigure[Convergence curves]{
        \includegraphics[scale=0.58]{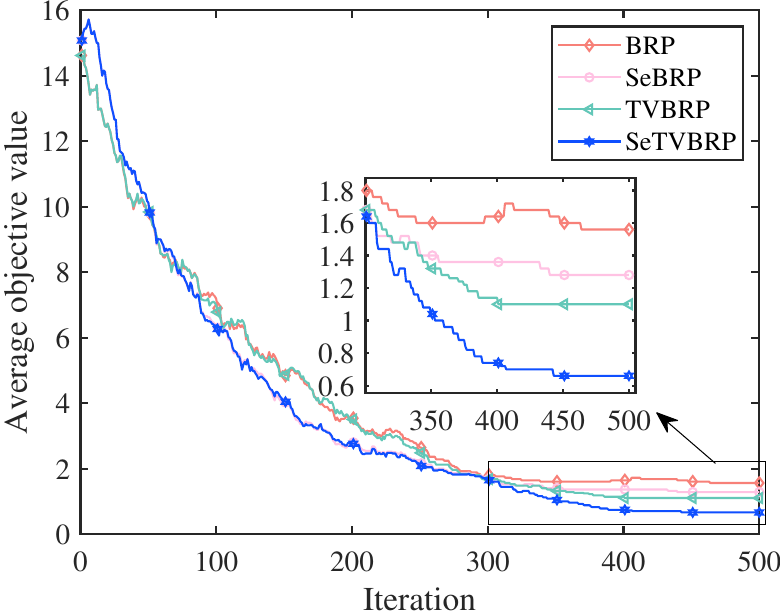}
    \label{f9a}
    }
    \quad
    \subfigure[Improvement in average objective value]{
        \includegraphics[scale=0.58]{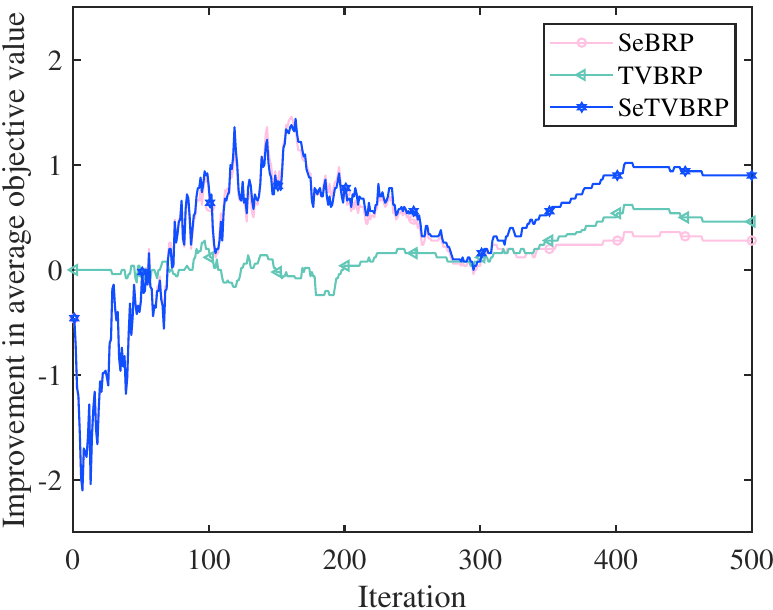}
    \label{f9b}
    }
\caption{Convergence performance of 4 algorithms for the regional scenario}
\label{f9}
\end{figure*}

We next investigate how the time-variant method and selective action method affect the performance of the SeTVBRP algorithm. Specifically, the proposed full algorithm (SeTVBRP algorithm) is compared with the original BRP algorithm, the algorithm with the time-variant method (TVBRP algorithm), and the algorithm with the selective action method (SeBRP algorithm).

The statistical results of these algorithms in the regional and global scenarios are given in Tables 3, where \emph{best} and \emph{worst} represent the best and worst cases of the system objective respectively, $\bar{R}$ denotes the system objective average, $T$ is the computation time, $\sigma $ is the variance, and ${{N}_{best}}$ is the number of runs in which the optimal solution is obtained out of 50 runs. The convergence curves of the different algorithms for the regional scenario are shown in Fig. \ref{f9}. Note that the global objective R represents the maximal residual workload of grids, and the solution improves with decreasing $R$. 

Table 3 shows that the proposed SeTVBRP algorithm achieves the best performance on each indicator in the regional scenario. This result illustrates the effect of the time-variant method and selective action method in terms of the solution quality and efficiency. On the one hand, the time-variant method improves the optimization ability of the algorithm; i.e., the average objective value $\bar{R}$ of the BRP algorithm is approximately 30\% less than that of the TVBRP algorithm (1.56 versus 1.04) and $\bar{R}$ of the SeBRP algorithm is approximately 48\% less than that of the SeTVBRP algorithm (1.28 versus 0.66). On the other hand, the selective action method helps reduce the computation time by shrinking the solution space for each iteration; i.e., the computation time T of the BRP algorithm is 35\% less than that of the SeBRP algorithm (1.225 versus 0.8064 s) and T of the TVBRP algorithm is 38\% less than the SeTVBRP algorithm (1.170 versus 0.6976 s). The convergence curves in Fig. \ref{f9} demonstrate that the algorithms with the selective action method (i.e., the SeBRP and SeTVBRP algorithms) preform better at 100–300 iterations and the algorithms with the time-variant method (i.e., the TVBRP and SeTVBRP algorithms) preform better at 300–500 iterations

\begin{table}[!ht]
\begin{center}
\setlength{\belowcaptionskip}{0.2cm}
\scriptsize
\caption{Allocation result of different algorithms}
\label{T2}
\begin{tabular}{lllllll}\hline
\multirow{2}{*}{Algorithms} & \multicolumn{6}{c}{Indications} \\ \cline{2-7} 
\multirow{2}{*}{}  & \emph{Worst} & \emph{Best} & $\bar{R}$ & $T$(s) & $\sigma$ & ${{N}_{best}}$\\ \hline
TVBRP & 2 & 0 & 1.04 & 1.169 & 0.937 & 22  \\ 
SeBRP & 4 & 0 & 1.28 & 0.806 & 1.530 & 21  \\
BRP & 4 & 0 & 1.56 & 1.225 & 1.190 & 14  \\ 
SeTVBRP & \textbf{2} & \textbf{0} & \textbf{0.66} & \textbf{0.697} & \textbf{0.841} & \textbf{32}  \\ \hline
\end{tabular}
\end{center}
\end{table}

\subsection{Algorithm comparison experiments}
\begin{figure}
	\centering
    \includegraphics[scale=0.59]{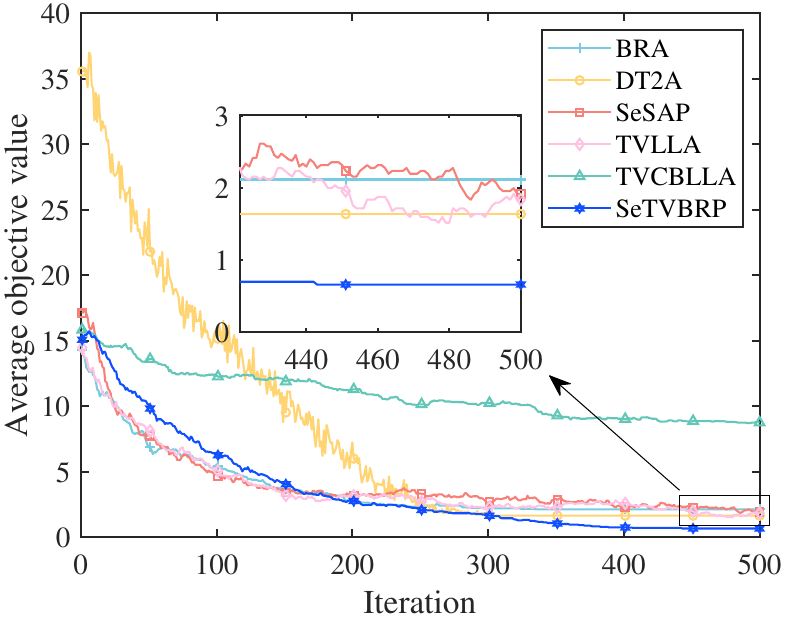}  
	\caption{Convergence performance of 6 algorithms for the regional scenario with 25 satellites and 9 grids for 500 iterations}   
\label{f10}
\end{figure}

To better present the superiority of the SeTVBRP algorithm in solving the sDGAP, we compare the algorithm with state-of-the-art potential game theory algorithms (i.e., BRA \cite{Ai.2008}, DT2A \cite{Sun.2021}, SeSAP \cite{GurdalArslan.2007}, TVLLA \cite{Sun.2018} and TVCBLLA \cite{Wu.2020}) in terms of the solution efficiency. The solution efficiency is reflected by two indicators, namely the value of the global objective and the CPU solution time. The comparison result of the regional scenario is given in Table 4, and the convergence curves of six algorithms are presented in Fig. \ref{f10}.  

The proposed SeTVBRP algorithm clearly performs best on all indicators and thus shows efficiency and robustness. On the one hand, as shown in Table 4, the SeTVBRP algorithm provides 32 optimal solutions over 50 runs within a computational time of only 0.7 s, whereas BRA, DT2A, SeSAP, and TVLLA provide fewer than 15 optimal solutions. On the other hand, the proposed SeTVBRP algorithm has the smallest variance among the six algorithms, and its value of the worst case solution is smallest, indicating the high robustness of the SeTVBRP algorithm and the good convergence of the algorithm under various initial settings. Although BRA takes little time to calculate a solution through the introduction of pure greediness, where agents only choose the best response, BRA performs the worst on the indicators relating to the solution quality owing to a lack of randomness. In contrast, because of the high randomness in the action selection, DT2A obtains a slightly better average solution than BRA (which is 1.64) but requires a massive computation time and has low robustness. TVCBLLA has difficulty reaching convergence by generation 500, and the solution is thus given when the number of iterations is 20,000, which is denoted TVCBLLA (20000). Compared with the SeTVBRP algorithm, TVCBLLA (20000) requires more iterations and takes a longer time to achieve convergence (i.e., approximately 10 times as long as the SeTVBRP algorithm). Its efficiency and robustness in solving the problem are inferior to those of the SeTVBRP algorithm. Furthermore, such an allocation algorithm requiring massive iterative communication between agents is not suitable for satellite systems, which incur a certain communication cost.

We next focus on the convergence curves of each learning method. Fig. \ref{f10} shows that the average objective value of the SeTVBRP algorithm is better than that of the other algorithms after 200 iterations. Before 200 iterations, the SeTVBRP algorithm performs slightly worse than BRA, SeSAP and TVLLA, owing to their commonly shared feature of strong greediness. In contrast, DT2A has more randomness in the rules of updating agent actions, which leads to poorer performance in the early evolution process but a better convergence result relative to the algorithms having strong greediness (i.e., BRA, SeSAP and TVLLA). The initial solution value obtained by DT2A markedly diverges from those of other algorithms. This is because that DT2A initiates with a random actions set in contrast to other algorithms that generate initial solution by actions with better objective values. \cite{Sun.2021} The proposed SeTVBRP algorithm lies somewhere in between, maintaining a balance of greediness and randomness so that it performs well in early iterations and obtains the best performance in the end among the algorithms.

\begin{table}[!ht]
\begin{center}
\setlength{\belowcaptionskip}{0.2cm}
\scriptsize
\caption{Allocation result of different state-of-the-art algorithms for the regional scenario}
\label{T3}
\begin{tabular}{lllllll}\hline
\multirow{2}{*}{Algorithms} & \multicolumn{6}{c}{Indications} \\ \cline{2-7} 
\multirow{2}{*}{} & \emph{Worst} & \emph{Best} & $\bar{R}$ & $T$(s) & $\sigma$ & ${{N}_{best}}$\\ \hline
SeTVBRP & \textbf{2} & \textbf{0} & \textbf{0.66} & \textbf{0.697} & \textbf{0.841} & \textbf{32}  \\ 
BRA & 8 & 0 & 2.12 & 0.884 & 3.822 & 11  \\ 
DT2A & 6 & 0 & 1.64 & 9.043 & 1.418 & 13  \\ 
SeSAP & 6 & 0 & 1.92 & 1.039 & 2.116 & 13  \\ 
TVLLA & 4 & 0 & 1.84 & 1.528  & 2.096  & 15  \\ \hline
TVCBLLA (20000) & 4 & 0 & 1.28 & 6.303 & 1.920 & 24  \\ 
CPLEX & - & - & 0 & 6.235 & - & -  \\ \hline
\end{tabular}
\end{center}
\end{table}

For the global scenario, we have comparison results qualitatively similar to those of the regional scenario. Note that the number of iterations for the results given in Table 4 has increased to 2000 because of the increases in the numbers of grids and satellites, and the time\-variant parameter $\tau$ value is set at 0.85. In this case, the proposed SeTVBRP algorithm still outperforms the other algorithms with an average objective value $\bar{R}=1.16$. Notably, the SeTVBRP algorithm obtains 17 optimal solutions over 50 runs whereas the remaining algorithms fail to find the optimal solution, which demonstrates the superiority of the SeTVBRP algorithm over the other algorithms in terms of the searching capacity on a large scale. 

\begin{figure}
	\centering
    \includegraphics[scale=0.59]{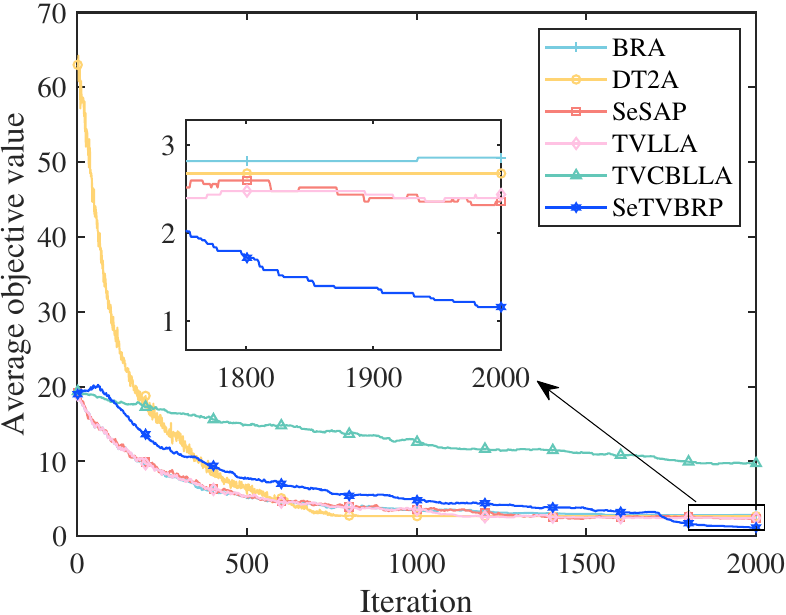}  
	\caption{Convergence performance of 6 algorithms for the global scenario with 100 satellites and 30 grids for 2000 iterations}   
\label{f11}
\end{figure}

Next, the convergence curves are considered. Fig. \ref{f11} shows a rapid drop in the average objective value for the SeTVBRP algorithm from 4 to 1.6 at iterations 1200–1400, and a similar trend is observed in the regional scenario at iterations 300–350. This is because the time-variant method of the SeTVBRP algorithm enters the exploitation phase at the last evolution. As in the regional scenario, the convergence curves of SeTVBRP start accelerating when iterations reach $\tau T^{max}$=1700 iterations. However, this does not mean that only when $\tau$=0.85 can SeTVBRP achieve better results than other algorithms. In fact, as shown in Fig. \ref{f6}, the SeTVBRP algorithm outperforms comparable algorithms within a range of $\tau$ from 0.2 to 0.95. In addition, the underwhelming performance of during the early convergence iterations does not imply a need for more computational resources. As evinced in Table 4, the computation of the SeTVBRP algorithm is the shortest among all algorithms, being only 4.628s. The reason is that the selective action method of the SeTVBRP algorithm helps an agent to sift out some actions to reduce the computation time with a slight sacrifice of the solution quality in early evolutions.

\begin{table}[!ht]
\begin{center}
\setlength{\belowcaptionskip}{0.2cm}
\scriptsize
\caption{Allocation result of different algorithms for the global scenario}
\label{T4}
\begin{tabular}{lllllll}\hline
\multirow{2}{*}{Algorithms} & \multicolumn{6}{c}{Indications} \\ \cline{2-7} 
\multirow{2}{*}{}  & \emph{Worst} & \emph{Best} & $\bar{R}$ & $T$(s) & $\sigma$ & ${{N}_{best}}$\\ \hline
SeTVBRP & \textbf{2} & \textbf{0} & \textbf{1.16} & \textbf{4.628} & \textbf{0.831} & \textbf{17}  \\ 
BRA & 6 & 2 & 2.86 & 10.042 & 1.715 & 0  \\ 
DT2A & 6 & 2 & 2.68 & 306.064 & 1.079 & 0  \\ 
SeSAP & 6 & 2 & 2.52 & 11.445 & 1.602 & 0  \\ 
TVLLA & 6 & 2 & 2.44 & 12.337 & 1.190 & 0  \\ \hline
TVCBLLA (20000) & 8 & 2 & 4.7 & 16.191 & 1.725 & 0  \\ 
CPLEX & - & - & 0 & 60.574 & - & -  \\ \hline
\end{tabular}
\end{center}
\end{table}

The SeTVBRP algorithm obtains average objective values of 0.66 and 1.16 in the two scenarios and approximates the optimal solutions as time goes to infinity. Moreover, the SeTVBRP algorithm is tested in more complex environments, where the number of satellites is increased to 200 and 300 respectively. We find that the proposed algorithms performs well in these complex scenarios. However, the performance of the SeTVBRP algorithm will drop if the granularity of the allocation time measurement decreases (e.g., from minutes to seconds). When the time granularity of allocation decreases, the size of the optional allocation set will increase exponentially, such that it is difficult for the agent to find the optimal solution in the huge action space.

\subsection{Multi-stage dynamic allocation experiments}

\begin{figure*}[!htb]
    \centering
    \subfigure[Grids observation load for multiple stage, the height of the pillars in the 3D histogram represents the observation load of grids, the $x$-axis represents the longitude index, where $x$=1 corresponds to the range of 90°E to 100°E; similarly, the $y$-axis represents the latitude index, where $y$=1 corresponds to the range of 0°N to 10°N]{
        \includegraphics[scale=0.48]{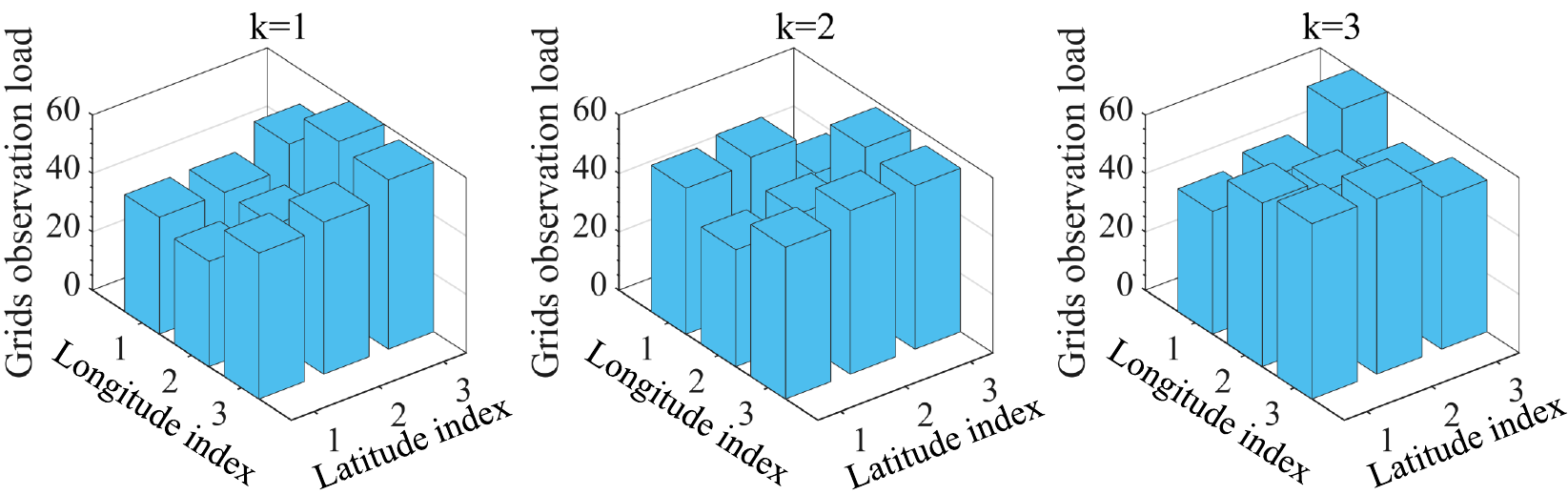}
    \label{F12A}
    }
    \quad
    \subfigure[Blue bold line represents the average convergence curves of SeTVBRP, and the light blue area represents the max-min range for each iteration over 50 runs, the pink dotted lines indicate the value of optimal solution]{
        \includegraphics[scale=0.38]{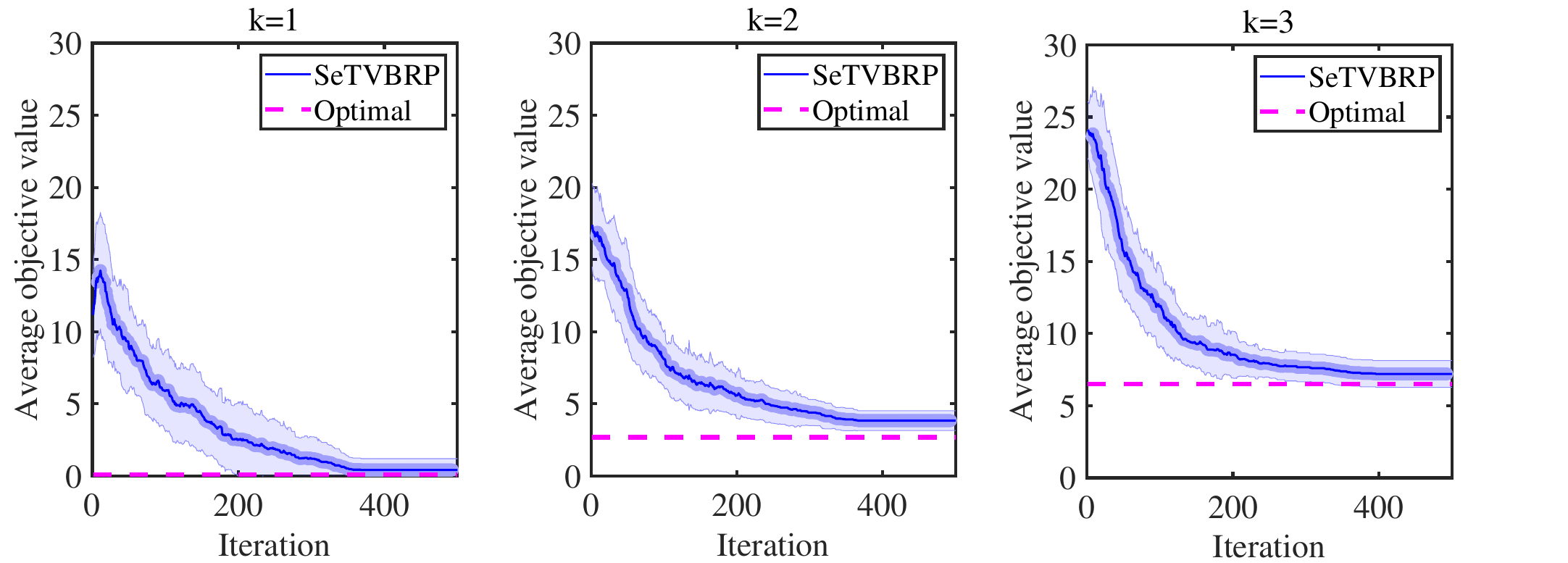}
    \label{F13B}
    }
\caption{the observation load of grids and performance of SeTVBRP for multiple stage}
\label{f12}
\end{figure*}

For the DGAP, only the regional scenario is considered, and the grid workloads and satellite work capacities are generated randomly as shown in Table 5 and Fig. \ref{f12}. The value of capacity and load constancy is based on a simplified model, which remains unchanged within a single stage over a small $\text{ }\!\!\Delta\!\!\text{ }{{t}}$ duration. Note that the transition time ${{\eta }_{ik}}$ at stage $k$ depends on the final allocation result at stage $k-1$, as shown by Eq. (2). Therefore, the optimal value obtained by the CPLEX optimizer changes according to the change in the allocation result of the previous stage, rather than being a fixed value. $\bar{R}$ of CPLEX presents the average optimal value under different initial conditions of the transition time ${{\eta }_{ik}}$. Additionally, to test the performance of the algorithm under observation overload, we increase the observation load as the stage number increases, as shown in Fig. \ref{f12}. Correspondingly, the CPLEX average optimal value increases. 

Table 5 presents the allocation results for multi-stage dynamic task allocation. This table shows that, with the constraints of the satellite transition time and the increase in observation load, the SeTVBRP algorithm achieves satisfactory performance in a short time (i.e., within 1.5 seconds) and obtains at least 13 optimal solutions in a total of 50 runs. The solving time of CPLEX increases exponentially with the observation workload.

Fig. \ref{f12}(a) shows the observation load of grids for the tested multiple stages. The overall observation load increases obviously with the stage number. Fig. \ref{f12}(b) presents the convergence curves of the SeTVBRP algorithm at different stages. The light blue area around the convergence curve in the figure represents the max–min range for each iteration over 50 runs. Under different initial solution settings, although the results of the algorithm have large ups and downs in the early iterations, satisfactory results are obtained within 400 iterations.

\begin{table}[!ht]
\begin{center}
\setlength{\belowcaptionskip}{0.2cm}
\scriptsize
\caption{Allocation result for multi-stage dynamic task allocation}
\label{T5}
\begin{tabular}{llllllll}\hline
\multirow{2}{*}{Stage $k$} & \multirow{2}{*}{Algorithms} & \multicolumn{6}{c}{Indications} \\ \cline{3-8} 
\multirow{2}{*}{}&\multirow{2}{*}{}& \emph{Worst} & \emph{Best} & $\bar{R}$ & $T$(s) & $\sigma$ & ${{N}_{best}}$\\ \hline
\multirow{2}{*}{1}&SeTVBRP & 2 & 0 & 0.42 & 0.679  & 0.616  & 38  \\
\multirow{2}{*}{ }&CPLEX & - & - & 0 & 2.372  & -  & - \\
 \multirow{2}{*}{2}& SeTVBRP & 6 & 2 & 3.62 & 1.389  & 0.771  & 13  \\
 \multirow{2}{*}{ }& CPLEX & - & - & 2.58 & 13.70  & -  & -  \\
 \multirow{2}{*}{3}& SeTVBRP & 8 & 6 & 7.2 & 1.319  & 0.857  & 13  \\ 
 \multirow{2}{*}{ }& CPLEX & - & - & 6.4 & 22.09  & -  & -  \\ \hline 
\end{tabular}
\end{center}
\end{table}

\section{Conclusion}
This paper investigated the DGAP by constructing a potential game theory framework to realize the cooperation of a distributed satellite system. The global goal of this problem is to minimize the maximum remaining observation load over all grids. We approximated this non-continuous minimax goal with a smooth global utility function though dual theory. A potential game theory framework was then established by decomposing the global function into the local utility of individuals. Additionally, we proposed the SeTVBRP algorithm as a distributed learning algorithm. Convergence to the Nash equilibrium within a finite number of iterations of the SeTVBRP algorithm was proved. 

We carried out parameter analysis experiments, algorithm comparison experiments and multi-stage dynamic allocation experiments. The parameter analysis experiments tested the performance of the SeTVBRP algorithm for different values of the time-variant method parameter $\tau$, which is the proportion of the exploration phase over all iterations. The results showed that the SeTVBRP algorithm performed better when the proportion of exploration ranged from 0.5 to 0.7. Moreover, the proportion of exploration increased slightly with the problem scale. Additionally, experiments on the effectiveness of two improvement methods (i.e., the time-variant method and selective action method) showed that the time-variant method improved the optimization ability of the algorithm, whereas the selective action method helped reduce the computation time by shrinking the solution space for each iteration. In comparison experiments, we compared our algorithm with state-of-the-art algorithms in two scenarios. The experiments showed that our algorithm was superior in terms of both the solution quality and the learning efficiency. Finally, we tested the performance of the algorithm under multi-stage dynamic allocation. The results showed that the SeTVBRP algorithm obtained better solutions in a shorter time than the CPLEX solver. 

Our future work will have two research directions. First, we plan to enhance the distributed learning algorithm with machine learning. The combination of the online distributed learning algorithm of game theory and machine learning is promising. Second, our future work will focus on the distributed satellite planning and scheduling toward the observation tasks in grids, with consideration of more realistic constraints and application. Specifically, in real-world situations, more complex satellite capability parameters with dynamic characteristics and time-dependent conversion time constraints, which are simplified in this paper, will be taken into consideration.

\large
\textbf{Acknowledgments}
\normalsize

This research was supported by the National Natural Science Foundation of China (72001212), the Young Elite Scientists Sponsorship Program by CAST (2022QNRC001) and the Hunan Provincial Innovation Foundation for Postgraduate (CX20200022). 

\bibliographystyle{ieeetr}
\bibliography{ref}

\newpage
\large
\textbf{Appendix}
\normalsize

To proof Claim 1, we will first state and prove Claim A.1.

\textbf{Claim A.1.} Consider the fragment linear minimization problem P0, the original objective function is as follows:

\begin{equation}
\text{minimize}\ \underset{j\in M}{\mathop{\max }}\,\left( {{\beta }_{j}}-\alpha _{j}^{T}x \right)
\label{eq35}
\end{equation}
The dual problem P3 is
\begin{equation}
\text{maximize}\ {{\beta }^{T}}\lambda 
\label{eq36}
\end{equation}
\begin{equation}
s.t.\quad {{A}^{T}}\lambda =0,{{1}^{T}}\lambda =1,\lambda \ge 0
\label{eq37}
\end{equation}

\textbf{Proof:} The original problem P0 can be equivalent to problem P2:
\begin{equation}
\text{minimize}\ \underset{j\in M}{\mathop{\max }}\,\left( {{y}_{j}} \right)
\label{eq38}
\end{equation}
\begin{equation}
s.t.\quad {{y}_{j}}={{\beta }_{j}}-\alpha _{j}^{T}x
\label{eq39}
\end{equation}
The dual function for problem P2 is
\begin{equation}
g\left( \lambda  \right)=\underset{x,y}{\mathop{\inf }}\,\left( \underset{j\in M}{\mathop{\max }}\,{{y}_{j}}+\underset{j=1}{\overset{m}{\mathop \sum }}\,{{\lambda }_{j}}\left( {{\beta}_{j}}-\alpha_{j}^{T}x-{{y}_{j}} \right) \right)
\label{eq40}
\end{equation}
The infimum over $x$ is finite only if $\sum\limits_{j\in M}{{{\lambda }_{j}}\alpha _{j}^{T}}={{A}^{T}}\lambda =0$, Thus, the original formula can be arranged as
\begin{equation}
g\left( \lambda  \right)=\underset{y}{\mathop{\inf }}\,\left( \underset{j\in M}{\mathop{\max }}\,{{y}_{j}}-\sum\limits_{j\in M}{{{\lambda }_{j}}{{y}_{j}}}+\sum\limits_{j\in M}{{{\lambda }_{j}}{{\beta }_{j}}} \right)
\label{eq41}
\end{equation}
Extract the part that is related to $y$:
\begin{equation}
\underset{y}{\mathop{\inf }}\,\left( \underset{j\in M}{\mathop{\max }}\,{{y}_{j}}-\sum\limits_{j\in M}{{{\lambda }_{j}}{{y}_{j}}} \right)
\label{eq42}
\end{equation}
(a) If $\lambda \ge 0\ {{1}^{T}}\lambda =1$, then $\sum\limits_{j\in M}{{{\lambda }_{j}}{{y}_{j}}}\le \sum\limits_{j\in M}{{{\lambda }_{j}}\underset{j}{\mathop{\max }}\,{{y}_{j}}=\underset{j}{\mathop{\max }}\,{{y}_{j}}}$ with equality if $y=0$, so in that case:
\begin{equation}
\underset{y}{\mathop{\inf }}\,\left( \underset{j\in M}{\mathop{\max }}\,{{y}_{j}}-\sum\limits_{j\in M}{{{\lambda }_{j}}{{y}_{j}}} \right)=0
\label{eq43}
\end{equation}
(b) If $\lambda <0$, selecting ${{y}_{j}}=0,i\ne j$ and $y_j=-t$ with $t>0$ gives
\begin{equation}
\underset{j\in M}{\mathop{\max }}\,{{y}_{j}}-\sum\limits_{j\in M}{{{\lambda }_{j}}{{y}_{j}}}=0-t{{\lambda }_{j}}\overset{t\to \infty }{\mathop{\to }}\,-\infty 
\label{eq44}
\end{equation}
(c) Finally, if ${{1}^{T}}\lambda \ne 1$, let $y=t\cdot 1$, then 
\begin{equation}
\underset{j\in M}{\mathop{\max }}\,{{y}_{j}}-\sum\limits_{j\in M}{{{\lambda }_{j}}{{y}_{j}}}=t{{\left( 1-{{1}^{T}}\lambda  \right)}^{t\to \infty ,{{1}^{T}}\lambda \ne 1}}\to -\infty 
\label{eq45}
\end{equation}
To sum up, CPL original formula can be arranged as
\begin{equation}
\underset{y}{\mathop{\inf }}\,\left( \underset{j\in M}{\mathop{\max }}\,{{y}_{j}}-\sum\limits_{j\in M}{{{\lambda }_{j}}{{y}_{j}}} \right)=\left\{ \begin{matrix}
   0,\ \lambda \ge 0,{{1}^{T}}\lambda =1  \\
   -\infty ,\ \text{otherwise}  \\
\end{matrix} \right.
\label{eq46}
\end{equation}
Therefore, the dual function of P2 is as follows
\begin{equation}
\begin{aligned}
  & g\left( \lambda  \right)=\underset{y}{\mathop{\inf }}\,\left( \underset{j\in M}{\mathop{\max }}\,{{y}_{j}}-\sum\limits_{j\in M}{{{\lambda }_{j}}{{y}_{j}}}+\sum\limits_{j\in M}{{{\lambda }_{j}}{{\beta }_{j}}} \right) \\ 
 & =\left\{ \begin{matrix}
   {{\beta }^{T}}\lambda \quad {{A}^{T}}\lambda =0\lambda \ge 0\ \ {{1}^{T}}\lambda =1  \\
   -\infty \quad \quad \quad \quad \quad \quad \ \quad \text{otherwise}  \\
\end{matrix} \right. \\ 
\end{aligned}
\label{eq47}
\end{equation}
The resulting dual problem P3 is
\begin{equation}
\text{maximize}\quad {{\beta }^{T}}\lambda 
\label{eq48}
\end{equation}
\begin{equation}
s.t.\quad {{A}^{T}}\lambda =0,{{1}^{T}}\lambda =1,\lambda \ge 0
\label{eq49}
\end{equation}
And that proves the claim.

\textbf{Claim 1.} Let $h\left( x \right)=\varepsilon \text{log}\left( \underset{j=1}{\overset{m}{\mathop \sum }}\,{{e}^{\frac{1}{\varepsilon }\left( {{\beta }_{j}}-\alpha _{j}^{T}x \right)}} \right)$. Suppose the problem P1 has the following objective function
\begin{equation}
\text{minimize}\quad h\left( x \right)
\label{eq50}
\end{equation}
Suppose the optimal value of P1 is $p_{1}^{*}$, then we have $0\le p_{1}^{*}-p_{0}^{*}\le ~\varepsilon \log m$.

\textbf{Proof:} Let ${{y}_{j}}={{\beta }_{j}}-\alpha _{j}^{T}x$, then the problem P1 is equivalent to problem P4:
\begin{equation}
\text{minimize}\quad \varepsilon \text{log}\left( \sum\limits_{j\in M}{{{e}^{\frac{1}{\varepsilon }{{y}_{j}}}}} \right)
\label{eq51}
\end{equation}
\begin{equation}
s.t.\quad {{y}_{j}}={{\beta }_{j}}-\alpha _{j}^{T}x
\label{eq52}
\end{equation}
then the Lagrangian function of P4 is
\begin{equation}
L\left( x,y,z \right)=\varepsilon \text{log}\left( \sum\limits_{j\in M}{{{e}^{\frac{1}{\varepsilon }{{y}_{j}}}}} \right)+{{z}^{T}}\left( \beta -Ax-y \right)
\label{eq53}
\end{equation}
, where $A=\left[ \alpha  \right]$, $L\left( x,y,z \right)$ is bounded below as a function of $x$ only if ${{z}^{T}}A=0$, to find the optimum over $y$, we set the gradient equal to zero.
\begin{equation}
\frac{\partial L\left( x,y,z \right)}{\partial y}=0
\label{eq54}
\end{equation}
That is
\begin{equation}
\frac{{{e}^{\frac{1}{\varepsilon }{{y}_{j}}}}}{\sum\limits_{j\in M}{{{e}^{\frac{1}{\varepsilon }{{y}_{j}}}}}}={{z}_{j}}
\label{eq55}
\end{equation}
This is solvable for $y_i$ if $z\ge 0,{{1}^{T}}z=1$. By substituting the above equation into the original one, we rewrite the Lagrangian function of P4 as
\begin{equation}
g\left( z \right)={{\beta }^{T}}z-\varepsilon \sum\limits_{j\in M}{{{z}_{j}}\log {{z}_{j}}}
\label{eq56}
\end{equation}
Overall, the dual problem P5 of the problem P4 is
\begin{equation}
\text{maximize}\quad {{\beta }^{T}}z-\varepsilon \sum\limits_{j\in M}{{{z}_{j}}\log {{z}_{j}}}
\label{eq57}
\end{equation}
\begin{equation}
s.t.\quad {{A}^{T}}z=0,{{1}^{T}}z=1,z\ge 0
\label{eq58}
\end{equation}
Suppose $z^{*}$ is the optimal solution for the dual problem P5, then $z^{*}$ is also feasible for the dual problem P3. Suppose the optimal value of P3 is $p_{3}^{*}$ and the optimal value of the original problem P0 is $p_{0}^{*}$, according to the Claim A.1, we have 
\begin{equation}
p_{0}^{*}=p_{3}^{*}\ge {{b}^{T}}{{z}^{*}}
\label{eq59}
\end{equation}
Then, since the problem P5 is the dual problem of P4, we have 
\begin{equation}
p_{5}^{*}=p_{4}^{*}={{b}^{T}}{{z}^{*}}-\varepsilon \sum\limits_{j\in M}{z_{j}^{*}\log z_{j}^{*}}
\label{eq60}
\end{equation}
, where $p_{4}^{*}$ and $p_{5}^{*}$ are the optimal value of problem P4 and P5. Combining the above two equations gives
\begin{equation}
p_{0}^{*}\ge p_{5}^{*}+\varepsilon \sum\limits_{j\in M}{z_{j}^{*}\log z_{j}^{*}}\ge p_{4}^{*}-\varepsilon \log m
\label{eq61}
\end{equation}
The bound follows from
\begin{equation}
\underset{{{1}^{T}}{{z}^{*}}=1}{\mathop{\min }}\,\sum\limits_{j\in M}{z_{j}^{*}\log z_{j}^{*}}=-\log m
\label{eq62}
\end{equation}
On the other hand, we have
\begin{equation}
\underset{j}{\mathop{\max }}\,\left( {{\beta }_{j}}-\alpha _{j}^{T}x \right)\le \varepsilon \text{log}\left( \sum\limits_{j\in M}{{{e}^{\frac{1}{\varepsilon }{{y}_{j}}}}} \right)
\label{eq63}
\end{equation}
for all $x$, and therefore $p_{0}^{*}\le p_{4}^{*}$. Since the problem P4 is equivalent to the problem P1, we have $p_{4}^{*}=p_{1}^{*}$. And this completes the proof.

\end{document}